\documentclass[aps,floats,preprint,showpacs,preprintnumbers,showkeys,floatfix]{revtex4}

\usepackage[latin1]{inputenc}                    
\usepackage{graphicx}                            
\usepackage{latexsym}                            
\usepackage{amsfonts}                            
\usepackage{amssymb}                             
\usepackage{amsmath}                             
\usepackage[mathscr]{eucal}                      
\usepackage{dcolumn}                             
\usepackage{multirow}                            
\usepackage{theorem}                             
\usepackage[normalem]{ulem}                      
\usepackage{psfrag}
\usepackage{rotating}
\usepackage{setspace}

\usepackage{fancyheadings}
\usepackage{graphicx} 
\usepackage{slashed}
\usepackage{amsmath}
\usepackage{amsfonts}
\usepackage{amssymb}

\usepackage{latexsym}                            
\usepackage[mathscr]{eucal}                      
\usepackage{dcolumn}                             
\usepackage{multirow}                            
\usepackage{theorem}                             
\usepackage[normalem]{ulem}                      
\usepackage{psfrag}

\newcommand{\bc}{\begin{center}}
\newcommand{\ec}{\end{center}}
\newcommand{\be}{\begin{equation}}
\newcommand{\ee}{\end{equation}}
\newcommand{\bea}{\begin{eqnarray}}
\newcommand{\eea}{\end{eqnarray}}

\newcommand{\emdash}{\hspace{1pt}---\hspace{1pt}}

\begin{document}

\preprint{
\vbox{
\hbox{ADP-13-17/T837}
}}

\title{Quark-Meson Coupling Model, Nuclear Matter Constraints \\
and Neutron Star Properties}
\author{D.~L.~Whittenbury}
\author{J.~D.~Carroll}
\author{A.~W.~Thomas}
\affiliation{CSSM and ARC Centre of Excellence for Particle Physics at the 
Terascale,\\ School of Chemistry and Physics,
University of Adelaide,
  Adelaide SA 5005, Australia}
  
\author{K.~Tsushima}
\affiliation{CSSM and ARC Centre of Excellence for Particle Physics at the 
Terascale,\\ School of Chemistry and Physics,
University of Adelaide,
  Adelaide SA 5005, Australia}
\affiliation{International Institute of Physics, Federal University of Rio Grande do Norte,
Natal-RN 59078-400, Brazil}
%

%
\author{J.~R.~Stone} 
\affiliation{Department of Physics, University of
  Oxford, Oxford OX13PU, United Kingdom}
\affiliation{Department of Physics and Astronomy, University of Tennessee, 
Knoxville TN 37996, USA}
\begin{abstract}
We explore the equation of state for nuclear matter in 
the quark-meson coupling 
model, including full Fock terms.  
The comparison with phenomenological 
constraints can be used to restrict the few additional parameters appearing 
in the Fock terms which are not present at Hartree level. Because the model 
is based upon the in-medium modification of the quark structure of the 
bound hadrons, it can be readily extended to include 
hyperons and to calculate the equation of state of dense matter in 
beta-equilibrium. This leads naturally to a study of the properties of 
neutron stars, including their maximum mass, their radii and density 
profiles.
\end{abstract}


\maketitle
\raggedbottom
%
\section{\label{sec:intro}Introduction}
Bulk nuclear matter properties have served as an excellent testing ground
for models of baryonic many-body systems for many years. This
hypothetical medium possesses many similarities with matter in the
interior of heavy nuclei, neutron stars and core-collapse
supernovae. The relative simplicity of the nuclear matter concept,
such as the assumption of a uniform density distribution without 
surface effects, allows the derivation of several key variables which
are generally accepted as necessary conditions that must be
satisfied by any successful nuclear model.

The uncertainty in the determination of the forces acting among
baryons and their modification by the medium has led to a great
variety of models. These traditionally start from a bare
nucleon-nucleon interaction, fit to experimental data from
nucleon-nucleon scattering and the properties of the deuteron, which
serves as input to a many-body formalism such as the relativistic
Dirac--Bruckner--Hartree--Fock (DBHF) approximation and its
nonrelativistic counterpart BHF~\cite{dbhf,bhf}, variational
methods~\cite{variational}, correlated basis function
models~\cite{cbf}, self-consistent Greens function (SCGF)
models~\cite{dewulf2003, frick2003}, quantum Monte Carlo
techniques~\cite{qmoca} and chiral effective field theory
\cite{hebeler2010a,hebeler2010b}. An alternative is to develop an
effective density dependent baryon-baryon interaction such as the
non-relativistic Skyrme or Gogny interaction, 
or one of the various relativistic
effective Lagrangian models and use it directly in a many-body
theory.

With the exception of the role of the $\Delta$ excitation in 
the generation of the three-nucleon force, none of these models 
consider the internal structure of the nucleon 
and, in particular, its possible modification in the presence of other 
hadrons. They depend on a large number of variable parameters which
are determined by fitting calculated observables to experimental data. 
The parameters are often
correlated, making it difficult to extract an unambiguous set from
such fits, leading to\emdash in principle\emdash an infinite number of
such parameter sets~\cite{Dutra:2012mb}.

The quark-meson coupling model (QMC model) is based upon a very different approach to this problem.
Rather than starting with the nucleon-nucleon (NN) force, it begins with 
the study of a hadron built from quarks immersed in a nuclear medium. The 
original model, which is employed here, begins with the MIT bag model. 
One then self-consistently includes the effects of the 
coupling to the $u$ and $d$ quarks of a 
scalar-isoscalar meson ($\sigma$) mean field, generated by all the 
other hadrons in the medium, on the internal structure of that hadron.
As in earlier boson-exchange models, the $\sigma$ is a crude but convenient 
way to simulate the effects of correlated two-pion exchange between hadrons.
While the quarks are also coupled to $\omega$ and $\rho$ mesons, their 
Lorentz vector nature means that, at least at Hartree level, they simply 
shift quark energies and do not generate non-trivial, density dependent 
modifications of the internal structure of the bound hadron.

The QMC model was originally introduced by
Guichon~\cite{Guichon:1987jp}. Subsequent development significantly improved 
the treatment of centre of mass corrections~\cite{Guichon:1995ue}, 
which had generated 
an unrealistic amount of repulsion in the original model. This development 
also included a consistent treatment of finite nuclei, including  
the spin-orbit force~\cite{Guichon:1995ue}. 
When applied to $\Lambda$ hypernuclei,  
the model provided a very natural explanation 
of the very small spin-orbit force 
observed in those 
systems~\cite{Tsushima:1997rd,Tsushima:1997cu,Guichon:2008zz}. 
In an important, recent development, the 
inclusion of the density dependence of the  ``hyperfine'' interaction 
between quarks arising from one-gluon-exchange (OGE) gave a parameter free 
explanation of the empirical absence of medium mass and heavy 
$\Sigma$-hypernuclei, while simultaneously yielding a good description 
of $\Lambda$-hypernuclei~\cite{Guichon:2008zz}. For a review of the 
many applications of the QMC model we refer to Ref.~\cite{Saito:2005rv}.

A clear connection has also been established between the self-consistent
treatment of in-medium hadron structure and the existence of
many-body~\cite{Guichon:2004xg} or density
dependent~\cite{Guichon:2006er} effective forces. {The Skyrme
interaction SQMC700, derived in \cite{Guichon:2006er}, was amongst
the few percent of Skyrme force which satisfied all the up-to-date
constraints on high density matter up to 3 times nuclear saturation
density recently examined by Dutra  {\it et
  al.}~\cite{Dutra:2012mb}.  In particular, in 
all of the models explored so far involving confined quarks, 
the self-consistent response to the applied mean scalar field tends to 
oppose that applied field. This effect can be represented as a ``scalar 
polarisability'' which effectively reduces the coupling of the $\sigma$ to 
an in-medium baryon as the applied scalar field increases. We stress that this 
scalar polarisability is a {\em calculated} property of each hadron and hence 
introduces no new parameters into the model. Moreover,  it is this scalar 
polarisability which yields the density dependence of the derived 
Skyrme forces, or equivalently the three-body forces between {\em all} 
combinations of hadrons. That is, the model predicts the existence and strength 
of the three-body forces between not just nucleons, but nucleons and hyperons 
and hyperons and other hyperons, without additional parameters. \\

As we have already observed, 
in a recent development of the QMC model~\cite{Guichon:2008zz}, the
self-consistent inclusion of the gluonic hyperfine interaction led to
a successful description of the binding energies of
$\Lambda$-hypernuclei, as well as the observed absence of medium and
heavy mass $\Sigma$-hypernuclei, with no additional parameters. 
We stress that these results were obtained under
the minimal assumption (consistent with the OZI rule) the $\sigma$,
$\omega$ and $\rho$ mesons do not couple to strange
quarks.

In this paper we present the latest development of the QMC model in
which we include the full vertex structure of the exchange term,
including not only the Dirac vector term, as was done in
~\cite{RikovskaStone:2006ta}, but also the Pauli tensor term. These
terms were already included within the QMC model by Krein~{\it et
  al.}~\cite{Krein:1998vc} for symmetric nuclear matter and more
recently by \cite{Miyatsu:2011bc}. We generalize the work of Krein~{\it et
  al.} by evaluating the full exchange terms for all octet baryons and
adding them, as additional contributions, to the energy density. A consequence
of this increased level of sophistication is that, if we insist on using the 
hyperon couplings predicted in the simple QMC model, with no coupling to
the strange quarks, the $\Lambda$ hyperon is no longer bound.

The present paper compliments the work of Ref.~\cite{Miyatsu:2011bc} who also considered the tensor interaction
in a variation of the QMC model,
by investigating an extended set of nuclear matter properties
with comparisons to heavy-ion collision data and other theoretical models.
 The present version of the QMC model differs
from \cite{Miyatsu:2011bc} as we use couplings as derived  within the model and treat
contact terms differently.  As is very well known from RMF \cite{walecka,Bouyssy:1987sh}
 and QMC\cite{Krein:1998vc} Hartree -- Fock
calculations the scalar $\Sigma^{\rm s}(k)$ and temporal vector $\Sigma^{0}(k)$ self-energy components are essentially independent of momentum and the spatial vector 
component is very small. For these reasons we make the assumption that the 
self-consistency can be treated approximately as in \cite
{RikovskaStone:2006ta} and as in \cite{Krein:1998vc} , where
the latter included a Fock correction to the scalar field. To state this more precisely 
we neglect the small spatial vector component of the self-energy such that 
$\vec{k}^{\ast} = \vec{k} + \hat{k}\Sigma^{v}(k) \simeq \vec{k}$ and the remaining
components are treated as momentum independent.
This approximate self-energy,
\begin{eqnarray}
\Sigma (k) & = & \Sigma^{\rm s}(k) -\gamma^{0}\Sigma^{0}(k) +\vec{\gamma}\cdot\vec{k}\Sigma^{\rm v}(k)\nonumber \\
&\simeq &\Sigma^{\rm s} - \gamma^{0}\Sigma^{0}
\end{eqnarray}
has identical form to the usual mean-field (Hartree) result and the Fock corrections to these components can be included by requiring thermodynamical consistency,
which amounts to minimising the total energy density with respect to the meson fields.
This results in a small correction to the scalar field.

In Sec.~\ref{sec:QMC} we present the basic features of the QMC model 
used in this work. The application of the model leading to the equation of 
state (EoS) of dense matter and a description of its parameters 
is given in Sec.~\ref{subsec:form}. Results obtained for infinite 
nuclear matter, symmetric and asymmetric as well as beta-equilibrium 
matter are followed by those for cold neutron stars and their comparison  
with experimental and observational constraints can be found in 
Secs.~\ref{subsec:snm} - \ref{subsec:cns}. The main results, sensitivity of the EoS and related quantities to variation of some model parameters is summarized 
in Sec.~\ref{subsec:sen}. We then make a comparison between 
the present work and recent variations of the QMC model studied 
in Refs.~\cite{Miyatsu:2011bc,Massot:2012pf} and others in
Sec.~\ref{subsec:comp}.
Discussion and concluding remarks are presented in Sec.~\ref{sec:dis}.

\section{\label{sec:QMC}The QMC model}

The QMC model is based upon the self-consistent modification of the
structure of a baryon embedded in nuclear matter. It is a relativistic
mean field model which incorporates the internal quark structure of
the baryons, represented as MIT bags containing three quarks 
in a color-singlet configuration. Interactions
occur between quarks in distinct bags via the exchange of mesons
coupled locally to the quarks.

Thus, in addition to the usual terms in the Lagrangian density of the 
MIT bag, the QMC model adds the simplest local couplings of 
$\sigma$, $\omega$ and $\rho$ mesons to the confined quarks. That is, 
the couplings are 
$g^{q}_\sigma \bar{q} q \sigma$, $g^{q}_\omega \bar{q} \gamma^\mu q \omega_\mu$ 
and $g^{q}_\rho \bar{q} \gamma^\mu \frac{\vec{\tau}}{2} q \cdot \vec{\rho}_\mu$, 
respectively~\cite{Guichon:1987jp,Guichon:1995ue}. 
Here $q$ represents the SU(2) doublet of $u$ and $d$ quarks 
and the coupling of these mesons to the $s$ quark is taken to be zero. 
These quark-meson couplings describe the interaction between quarks in 
{\em different} hadrons. They act as the source of mean fields in-medium 
as well as serving to modify the equation of motion of the confined quarks. 
This leads to a self-consistency problem which is highly non-trivial for 
the scalar field, whereas the vector couplings in uniform, infinite nuclear matter 
involve only time components -- e.g., $\omega^\mu = 
\bar{\omega} \delta^{\mu 0}$ -- and so they simply shift energy levels.
As a result, the effective strength of the coupling of the scalar meson 
to a hadron containing light quarks is suppressed as the scalar field 
increases \emdash or equivalently, as the density increases.
Thus, as a result of this self-consistent calculation at the 
quark level, one can express the 
in-medium baryon masses, $M^{\ast}_{B}$, 
as functions of the scalar field (as in
Ref.~\cite{Guichon:2008zz}) through a calculated, density dependent, 
scalar coupling,
$g_{\sigma B}(\bar{\sigma})$.

The saturation of symmetric nuclear matter~\cite{Guichon:1987jp} is a
natural effect of the self-consistent response of the quark wave
functions to the mean scalar field, a direct consequence of which is
the reduction of the effective $\sigma N$ coupling as the
$\sigma$-field increases. By analogy with the electric polarisability 
of an atom, which tends to arrange its internal structure to oppose an 
applied electric field, this reduction of the $\sigma N$ coupling 
is characterised as the scalar
polarisability of the nucleon. It is remarkable that the  
influence of baryon
sub-structure, in a mean field approximation, is entirely described 
in terms of the parameterisation of the effective mass of the baryon 
through the density dependent scalar coupling derived from the quark 
model of the baryon and $g^{q}_\sigma$. One can therefore replace the 
explicit description of the internal structure of the baryons by 
constructing an effective Lagrangian on the hadronic level, with 
the calculated non-linear $\sigma$-baryon couplings
given in~\cite{Guichon:2008zz}
\begin{equation}
M^*_B = M_B - w_{\sigma B}g_{\sigma N}\bar{\sigma}
+ \frac{d}{2}\tilde{w}_{\sigma B}\left(g_{\sigma N}\bar{\sigma}\right)^2 \ ,
\end{equation}
(where the weightings $w_{\sigma B}$ and $\tilde{w}_{\sigma B}$ simply
allow the use of a unique coupling to nucleons)
and proceed
to solve the relativistic mean field equations in a standard
way~\cite{walecka}.

The QMC Lagrangian density used in this work is given by a combination
of baryon, meson, and lepton components
\begin{equation}
\mathcal{L} = \sum_{B}\mathcal{L}_{B} 
+ \sum_m \mathcal{L}_{m}
+ \sum_\ell \mathcal{L}_\ell\ , 
\end{equation}
for the octet of baryons \mbox{$B \in
  \{N,\Lambda,\Sigma,\Xi\}$}, selected
mesons \mbox{$m \in \{\sigma,\omega,\rho,\pi\}$}, and leptons
\mbox{$\ell \in \{e^-,\mu^-\}$} with the individual Lagrangian
densities
\begin{equation}
\mathcal{L}_{B} =
\bar{\Psi}_{B}\left(
i\gamma_{\mu}\partial^{\mu} - M_{B} + g_{\sigma B}(\sigma)\sigma
- \Gamma^{\mu}_{\omega B}\omega_{\mu}
- \vec{\Gamma}_{\rho B}^{\mu} \cdot\vec{\rho}_{\mu} 
- \vec{\Gamma}_{\pi B}\cdot\vec{\pi}
\right) \Psi_{B}\ ,
\label{eq:lb}
\end{equation}
\begin{eqnarray}
\sum_{m}\mathcal{L}_{m} &=& \frac{1}{2}(\partial_{\mu}\sigma\partial^{\mu}\sigma 
- m_{\sigma}^{2}\sigma^{2}) 
- \frac{1}{4}\Omega_{\mu\nu}\Omega^{\mu\nu} 
+ \frac{1}{2}m_{\omega}^{2}\omega_{\mu}\omega^{\mu} \nonumber \\
&& 
- \frac{1}{4}\vec{R}_{\mu\nu}\cdot\vec{R}^{\mu\nu} 
+ \frac{1}{2}m_{\rho}^{2}\vec{\rho}_{\mu}\cdot\vec{\rho}^{\mu}
+ \frac{1}{2}(\partial_{\mu}\vec{\pi}\cdot\partial^{\mu}\vec{\pi} 
- m_{\pi}^{2}\vec{\pi}\cdot \vec{\pi})\ ,
\end{eqnarray}
for which the vector meson field strength tensors are 
$\Omega_{\mu\nu}=\partial_{\mu}\omega_{\nu}-\partial_{\nu}\omega_{\mu}$
and
$\vec{R}_{\mu\nu}=\partial_{\mu}\vec{\rho}_{\nu}-\partial_{\nu}\vec{\rho}_{\mu}$, and
\begin{equation}
{\mathcal L}_\ell =
\bar{\Psi}_\ell\left(i\gamma_\mu\partial^\mu-m_\ell\right)\Psi_\ell\ .
\end{equation}
For the baryon masses we take the average over the isospin multiplet of
their experimental values, where as for the mesons and leptons we simply
use the experimental values.

In a mean-field description of infinite nuclear matter with uniform 
density, one can set spatial derivatives of all fields to zero and replace 
the meson field operators by their expectation values:
\begin{eqnarray}
\sigma &\to& \langle\sigma\rangle \equiv \bar{\sigma}\ , \\
\omega_{\mu} &\to& \langle\omega_{\mu}\rangle 
= \langle \delta_{\mu 0}\omega_{\mu}\rangle \equiv \bar{\omega}\ , \\
\vec{\rho}_{\mu} &\to& \langle\vec{\rho}_{\mu}\rangle 
= \langle \delta_{\mu 0}\delta_{a 3}\rho_{\mu a}\rangle 
\equiv \bar{\rho}\ , \\
\vec{\pi} &\to & \langle \vec{\pi} \rangle = 0\ .
\end{eqnarray}
This is usually called the Hartree mean-field approximation.

The next step is to include the Fock level contributions  
involving the meson baryon vertices
which are expressed as

\begin{eqnarray}
\label{eq:sigmaGamma}
\Gamma_{\sigma B} &=& 
g_{\sigma  B}C_{B}(\bar{\sigma})F^{\sigma}(k^2)\mathbf{1}
= -\frac{\partial M^{\ast}_{B}}{\partial \bar{\sigma}}F^{\sigma}(k^2)\mathbf{1}\ , \\[2mm]
\label{eq:vectorGamma}
\vec{\Gamma}_{\eta B} &=& \epsilon^{\mu}_{\eta}\vec{\Gamma}_{\mu\eta B} 
= \epsilon^{\mu}_{\eta}\left[g_{\eta B} \gamma_{\mu} F^{\eta}_{1}(k^{2}) 
+ \frac{i f_{\eta B}\sigma_{\mu\nu}}
{2M^{\ast}_{B}}k^{\nu}F^{\eta}_{2}(k^{2})\right]\, \vec{t}\ ;\ 
\eta \in \{\omega, \rho\}\ ,\\
\label{eq:piGamma}
\vec{\Gamma}_{\pi BB'} &=& ig_{\pi BB'} F^{\pi}(k^{2})
\gamma^{\mu}k_{\mu} \gamma_{5}\vec{\tau}\ ,
\end{eqnarray}
with the isospin matrix $\vec{t}$ only applicable to isovector mesons.
For nucleons and cascade particles $\vec{t} = \frac{\vec{\tau}}{2}$. 
For the rho meson the flavour dependence is contained completely 
in the isospin matrix, such that $g_{\rho B} =g_{\rho N}=g_{\rho}$.
The pion-baryon interaction is assumed to be described by an
SU(3) invariant Lagrangian with the mixing parameter $\alpha = 2/5$ \cite{RikovskaStone:2006ta}
from which the hyperon -- pion coupling constants can be given in terms of
the pion nucleon coupling,  $g_{\pi BB'} = g_{\pi NN} \chi_{BB'}=\frac{g_{A}}{2f_{\pi}}\chi_{BB'}$
\cite{deSwart:CGC,RikovskaStone:2006ta}.

The ratios of tensor to vector couplings $\kappa^{B}_{(\omega,\rho)} =
f_{B(\omega,\rho)}/g_{B(\omega,\rho)}$ given in
Table~\ref{table:kappas} are rescaled using the free proton mass
\begin{equation}
\kappa^{B}_{(\omega,\rho)} \to \kappa^{B}_{(\omega,\rho)} \times\frac{M^{\ast}_{B}}{M_{p}} \  .
\label{eq:kapRescale}
\end{equation}
Equation (\ref{eq:kapRescale}) is used in all variants of the model (``scenarios"), considered in this work 
 except where a result is labeled ``Eff. Proton Mass".
The reason for this choice is that the derivation of the QMC model is based on an order by order expansion in the effect of the scalar field, using the effective mass of the proton in the Pauli term coupling assumes that the scalar field does not appear in some other way at the level of momentum dependent couplings. A systematic expansion would ensure that all effects are included consistently to a given order. In the absence of such a derivation it would be natural to write the couplings in terms of the free baryon mass as is done in Ref.~\cite{Miyatsu:2012xh,Miyatsu:2011bc} and not include just one effect of the scalar field at this order. 

The $\sigma$, $\omega$, $\rho$ and $\pi$ form factors are all taken to have
the dipole form $F(k^{2}) \simeq F(\vec{k}^{2})$ with the same
cutoff $\Lambda$. Clearly, these form factors are only of concern for
the Fock terms. We make specific note of the two terms which
contribute to the vector meson vertices, a vector `Dirac' term and a
tensor `Pauli' term. 

Through the Euler-Lagrange equations, we obtain from this Lagrangian
density a standard system of coupled, non-linear partial differential
equations for the meson mean fields~\cite{Guichon:1995ue}.
Meson retardation effects are not included and contact terms are
subtracted -- see the Appendix for details. We note that the mean
field approximation becomes progressively more reliable with
increasing density. Finally, we note that we have also neglected 
any modification of the Dirac sea of negative energy states 
with increasing density (see, however,
the discussion of such effects within the NJL model in Ref.~\cite{Bentz:2001vc}).

\section{\label{sec:EoS}Equation of State of baryonic matter}
\subsection{\label{subsec:form} Formalism}

The equation of state relates energy density, pressure, and 
temperature to baryon
number densities $\rho_{B}$. In this work, we include contributions from 
the full baryon octet in the limit $T = 0$.
The total energy density is given as a sum of the baryonic, 
mesonic and leptonic contributions 
\begin{equation}
\epsilon_{\rm total} = 
\epsilon_{B} + \epsilon_{\sigma \omega \rho} + \epsilon_{F} + 
\epsilon_{\ell} \, .
\label{eq:total} 
\end{equation}
The non-leptonic energy density can be divided into a
direct (Hartree) part, $\epsilon_{H} = \epsilon_{B} +
\epsilon_{\sigma\omega\rho}$, where
\begin{eqnarray}
\epsilon_{B} &=& \frac{2}{(2\pi)^{3}}\sum_{B} 
\int\limits_{\vert \mathbf{p}\vert <p_{F}}d\mathbf{p}
\sqrt{p^{2}+M^{\ast\, 2}_{B}} \ , \\
\epsilon_{\sigma\omega\rho} &=&
\sum_{\alpha=\sigma,\omega,\rho}
\frac{1}{2}m^{2}_{\alpha}\bar{\alpha}^{2}\ 
\label{eq:mean}, 
\end{eqnarray}
where $\bar{\alpha}$ refers to the mean field value of meson $\alpha$, 
plus an exchange (Fock) contribution 
\begin{equation}
\label{eq:fock}
\epsilon_{F} = \frac{1}{(2\pi)^{6}}\sum_{m=\sigma,\omega,\rho,\pi}\sum_{BB'} 
C^{m}_{BB'} \int\limits_{\substack{\vert \mathbf{p}\vert <p_{F}}}
\int\limits_{\substack{\vert \mathbf{p}'\vert <p_{F'}}} d\mathbf{p} d\mathbf{p}' \ 
\mathbf{\Xi}_{BB'}^{m}\ .
\end{equation}
The coefficients
$C^{\sigma}_{BB'}=C^{\omega}_{BB'}=\delta_{BB'}$, $C^{\rho}_{BB'}$,
and $C^{\pi}_{BB'}$, which arise from symmetry considerations, are
given in Ref.~\cite{RikovskaStone:2006ta}. This non-leptonic energy
density is then given by $\epsilon_{\rm hadronic} = \epsilon_H +
\epsilon_F = \epsilon_B + \epsilon_{\sigma \omega \rho} + \epsilon_F$.
Note that the pion contributes only at exchange level as parity
considerations lead to a vanishing direct level
contribution. It nonetheless plays an
important role in reducing the incompressibility of nuclear 
matter~\cite{RikovskaStone:2006ta}.

The leptonic energy density is simply
\begin{equation}
\epsilon_{\ell} = \frac{2}{(2\pi)^{3}}\sum_{\ell} 
\int\limits_{\vert \mathbf{p}\vert <p_{F,\ell}}d\mathbf{p}
\sqrt{p^{2}+m^2_{\ell}}\ .
\end{equation}

The scalar mean field in Eq.~(\ref{eq:mean}) is calculated self-consistently as
\begin{eqnarray}
\bar{\sigma} &=&
- \frac{1}{m^{2}_{\sigma}}\frac{\partial \epsilon_{H}}{\partial\bar{\sigma}}
- \frac{1}{m^{2}_{\sigma}}\frac{\partial \epsilon_{F}}{\partial \bar{\sigma}} \\
&=& 
- \frac{2}{m^{2}_{\sigma}(2\pi)^{3}}
\sum_{B}\int\limits_{\vert \mathbf{p}\vert <p_{F}} d\mathbf{p}
\frac{M^{\ast}_{B}}{\sqrt{p^{2}+M^{\ast\, 2}_{B}}}
\frac{\partial M^{\ast}_{B}}{\partial\bar{\sigma}}
- \frac{1}{m^{2}_{\sigma}}\frac{\partial \epsilon_{F}}{\partial \bar{\sigma}} \ ,
\label{eq:sigmean}
\end{eqnarray} 
where the second term in Eq.~(\ref{eq:sigmean}) is the Fock level correction to the scalar field, which
is included in the scenarios ``Fock $\delta\sigma$'' and ``Eff. Proton Mass $+ \ \delta\sigma$".
The vector meson mean fields simply scale with 
either the total or isovector baryonic density
\begin{eqnarray}
\bar{\omega} &=& \sum_B \frac{g_{\omega B}}{m^{2}_{\omega}} \rho_{B} \ ,\\
\bar{\rho} &=& \sum_B \frac{g_{\rho B}}{m^{2}_{\rho}}I_{3B} \rho_{B} \ ,
\end{eqnarray}
where $I_{3B}$ is the third component of isospin for baryon $B$.

For $\epsilon_{F}$, shown in Eq.~(\ref{eq:fock}), the integrand has the form 
\begin{equation}
\mathbf{\Xi}^{m}_{BB'}  = \frac{1}{2}\sum_{s,s'} \vert
\bar{u}_{B'}(p',s')\Gamma_{m BB'} u_{B}(p,s)\vert^{2} 
\Delta_{m}(\mathbf{k})\ ,
\end{equation}
where $\Delta_{m}(\mathbf{k})$ is the Yukawa propagator for meson
$m$ with momentum $\mathbf{k} =\mathbf{p} - \mathbf{p'}$, and
$u_{B}$ are the baryon spinors. The integrands are presented in the
Appendix.

The expression for total energy density is therefore dependent on  
just the three main adjustable coupling constants, which control 
the coupling of 
the mesons to the two lightest quarks, $g^{q}_{\sigma}$,
$g^{q}_{\omega}$, and $g^{q}_{\rho}$ for $q=u,d$ 
($g^{s}_{\alpha}=0$ for all mesons $\alpha$). In addition, one has the
meson masses, the value of the cut-off parameter 
$\Lambda$ appearing in the dipole form factors needed to evaluate 
the Fock terms and finally 
the bag radius of the free nucleon.
The $\sigma$, $\omega$, and $\rho$ couplings to the quarks are
constrained to reproduce the standard empirical properties of 
symmetric ($N$=$Z$) nuclear matter; the saturation 
density $\rho_{\rm 0}$ = 0.16 fm$^{\rm -3}$, the 
binding energy per nucleon at saturation 
of $\mathcal{E}(\rho = \rho_{0}) = -15.865$~MeV 
as well as 
the asymmetry energy coefficient \mbox{$a_{\rm asym} \equiv
  S_{0} \equiv S(\rho_{0}) = 32.5$~MeV}~\cite{RikovskaStone:2006ta} 
(see also Secs.~\ref{subsec:anm}). 

The $\omega$, $\rho$ and $\pi$ meson masses are set to 
their experimental values.  The
ambiguity in defining the mass of the $\sigma$ after quantising the
classical equations of motion was explained in detail in
Ref.~\cite{Guichon:1995ue}. Here it is set to the value that gave the
best agreement with experiment for the binding energies of finite nuclei
in a previous QMC model
calculation~\cite{Guichon:2006er}, which was $700$~MeV. This is a common
value taken for the sigma meson mass which is generally considered in
RMF models to be in the range $400$--$800$~MeV.

The form factor cut-off mass, $\Lambda$, controls the
strength of the Fock terms Eqs.~(\ref{eq:sigmaGamma} - \ref{eq:piGamma}). 
We considered a range of values; $0.9~{\rm
  GeV} \leq \Lambda \leq 2.0~{\rm GeV}$, with the preferred value, 
as we shall see, being 0.9 GeV.
For simplicity we have used the same
cutoff for all mesons. Since the pion mass is much lower than that of
the other mesons, we have confirmed that using a lower cutoff for the
pion does not significantly influence the results. This is not surprising as
Fock terms are expected to be more significant at higher density where
we have found that the pion does not contribute greatly.

All the other coupling constants in the expression for the total energy 
density are \textit{calculated} within the QMC model or determined from 
symmetry considerations without further need for adjustable parameters. 
The one exception is $g_{\sigma B}(\bar{\sigma})$, which shows a weak 
dependence on the free nucleon radius $R_{N}^{\textrm{free}}$. 
We checked that changes of order 20\% in
$R_{N}^{\textrm{free}}$, consistent with nucleon properties, 
have no significant effect on the properties of nuclear matter and 
chose $R_{N}^{\mathrm{free}}=1.0$~fm.

The baryon-meson coupling constants $g_{\sigma N}(0)$, 
$g_{\omega B}$, and $g_{\rho B}$ (or equivalently the three quark-meson 
coupling constants) are determined by
fitting the saturation properties of symmetric nuclear matter. 
Only $g_{\sigma B}$ is density dependent and that dependence is 
calculated self-consistently according to

\begin{equation}
\frac{\partial}{\partial\bar{\sigma}}\left[ 
g_{\sigma B}(\bar{\sigma})\bar{\sigma}\right] 
= g_{\sigma B}(0) \, C_{B}(\bar{\sigma}) = - \, \frac{\partial
  M^{\ast}_{B}}{\partial \bar{\sigma}} \equiv - \, \frac{\partial
  M^{\ast}_{B}(\bar{\sigma},g_{\sigma N},R^{\mathrm{free}}_{N})}{\partial
  \bar{\sigma}} \ ,
  \label{eq:vectcc}
\end{equation}
where $M^{\ast}_{B}$ is calculated in the QMC model using the MIT bag
with one gluon exchange for the baryon structure.
The couplings $g_{\omega B}$ and $g_{\rho B}$ are expressed in terms of 
the quark level couplings as:
\begin{equation}
g_{\omega B}=n^{B}_{u,d}g^{q}_{\omega} \,\, ; \, \,
g_{\rho B} = g_{\rho N} = g^{q}_{\rho}\ ,
\end{equation}
where $n^{B}_{u,d}$ is the number of light quarks in baryon $B$.

At densities $\sim$ 2 -- 3 $\rho_{\rm 0}$  
one expects, simply because the Fermi level of the neutrons rises rapidly,  
that for matter in beta-equilibrium hyperons must be considered. 
There is very little experimental data on the $N$--$Y$ and $Y$--$Y$ 
interactions, which makes the traditional approach through phenomenological 
pair-wise interactions very difficult. There is certainly no hope of 
determining the relevant three-body forces which are expected to be 
critical at high density. One of the attractive features of the 
QMC model is that it predicts all of these forces in terms of the 
underlying quark-meson couplings, the scalar meson mass and the 
particular quark model chosen (the MIT bag here).
Furthermore, the density dependence  of the scalar couplings to 
each baryon is also 
determined by the bag model mass parameterisation. 
The inclusion of this density dependent,  
in-medium interaction is equivalent in a density 
independent framework to including the appropriate three-body force 
between all baryons.

Remarkably, in the absence of the Pauli Fock terms, the model predicted realistic $\Lambda$ binding energies and, at
the same time realistic $\Sigma$ repulsion in matter
\cite{Guichon:2008zz}) . As we show later in
Sec.~\ref{subsec:sen}, the additional repulsion associated with the Fock
term, is not adequately compensated and the
agreement is lost. In this work we assess the magnitude of the needed
change by artificially modifying the $\sigma$--couplings for the hyperons to match the
empirical observations. This procedure will serve as a guidance in the
future development of the model, as outlined in Sec.~\ref{sec:dis}

It is well known that the coupling of the $\rho$ meson to a 
particular baryon  
has a relatively large Pauli, or tensor, 
coupling (i.e. $f_{\rho B}$ in Eq.~(\ref{eq:vectorGamma})). 
The value used varies 
from one model of the nuclear force to another. In the QMC model 
the prediction of the 
tensor coupling at zero momentum transfer is unambiguous \emdash
it is exactly the anomalous, iso-vector 
magnetic moment of the baryon in the MIT bag model.
Similarly, the tensor coupling of the $\omega$, which in the case
of the nucleon is much smaller than for the $\rho$, is determined by 
the isoscalar magnetic moment.
Since the MIT bag model reproduces the experimental values 
of the magnetic moments quite well, 
the tensor coupling required 
within the QMC model is equivalent to using vector meson 
dominance~\cite{Williams:1996id} and in practice 
we use values for the magnetic moments
from the Particle Data Group~\cite{Beringer:1900zz}.
Finally and purely as an exercise aimed at exploring
the model dependence, we consider two different choices for the ratios of
tensor to vector coupling constants 
${\displaystyle f_{\alpha B}/g_{\alpha B}}$; 
with $\alpha \in \{\rho,\omega\}$. 
Whereas, as we explained, in the standard QMC calculation we take 
\mbox{${\displaystyle f_{\rho  N}/g_{\rho N}} = 3.70$},
we also explore the
consequences of arbitrarily setting  
\mbox{${\displaystyle f_{\rho N}/g_{\rho N}} = 5.68$} in the
  `Increased $f_{\rho N}/g_{\rho N}$' scenario. In this scenario we arbitrarily 
 take the ratios of tensor to vector couplings of all baryons from 
 the Nijmegen potentials (Table VII of Ref. \cite{Rijken:2010zzb}).

The only other parameters in the QMC model 
are those entering the bag model.
We refer the reader to
Ref.~\cite{Guichon:2008zz} where those parameters were
obtained. None of them have
been adjusted to any property of nuclear matter, although all
calculations involving the QMC model at present rely on the MIT bag model
with one gluon exchange and could be in principle improved upon by
using a more sophisticated model of quark confinement. Nonetheless,
with this simple quark-based model, remarkable agreement with
a broad range of experimental data 
has been obtained~\cite{Saito:2005rv}.

Having established the QMC model parameters, in the following section we 
calculate properties of symmetric (SNM) and pure neutron (PNM) 
nuclear matter as well as matter in beta-equilibrium (BEM).
The latter consists of nucleons and leptons, while matter in generalized 
beta-equilibrium (GBEM) contains the full baryon octet and leptons. 
Using the derived EoS, we calculate the properties of 
cold neutron stars and make a comparison with up-to-date experimental 
and observational data. We also examine the robustness of those 
results on the limited number of parameters entering the model.\\
\subsection{\label{subsec:snm}Infinite symmetric and pure neutron nuclear matter}
A minimal set of saturation properties of symmetric nuclear matter, 
the saturation density, the binding energy per particle and the 
symmetry energy at saturation, were used to fix the quark-meson coupling 
constants as described in Sec.~\ref{subsec:form}. None of 
those properties is actually an \textit{empirical} quantity, since they 
are not measured directly but extracted from experiments or observations 
in a model dependent way. However, there is a general consensus that 
all meaningful theories of nuclear matter should reproduce these 
quantities correctly. Moreover, other properties of both symmetric 
and pure neutron matter, derived from derivatives of the energy 
per particle with respect to particle number density, 
together with their density dependence, can be compared to empirical 
data to further test the theories. These include the pressure, 
incompressibility (compression modulus) and the slope of 
the symmetry energy.    

Let us define the hadronic energy per particle,
$E = \epsilon_{\rm hadronic}$/$\rho$, where $\rho$ is the 
total baryonic density and define the following quantities 
as a function of $\rho$:
The first derivative of $E$ provides an expression for baryonic pressure
\begin{equation}
P=\rho^{2} \frac{\partial{E}}{\partial\rho} \, .
\end{equation}
The second derivative of $E$ is the compression modulus or incompressibility
\begin{equation}
K=9\rho^{\rm 2}\left(\frac{\partial^{\rm 2}E}{\partial{\rho^{\rm 2}}}\right) \, .
\end{equation}
The third derivative defines the so called skewness coefficient 
(some authors define $K' = -Q$)
\begin{equation}
Q=27\rho^{\rm 3}\left(\frac{\partial^{\rm 3}E}{\partial{\rho^{\rm 3}}}\right).
\end{equation}
These quantities can be evaluated at any density and any proton/neutron 
asymmetry ratio $\beta = (\rho_{\rm n} - \rho_{\rm p})/\rho$ at 
which the model for the baryonic energy per particle is valid.
The particular values at saturation density, $\rho_0$, are indicated 
with a subscript zero (e.g., $K_0, \, Q_0$ etc.). 
In symmetric nuclear matter, $\rho_{\rm n} = \rho_{\rm p}$ = 1/2 $\rho$, 
the values of the incompressibility and skewness at saturation density 
can be compared 
to experiment. Obviously, the pressure at saturation density 
is equal to zero. It is convenient to express the density 
dependence of the energy per particle in SNM as a Taylor expansion 
of $E$ about the saturation density in terms of a variable 
$x = (\rho - \rho_{\rm 0})/3\rho_{\rm 0}$
\begin{equation}
E_{\rm SNM}(\rho)=E_{\rm 0} +\frac{1}{2}K_{\rm 0}x^{\rm 2}+
\frac{1}{6}Q_{\rm 0}x^{\rm 3} + \mathcal{O}(x^4) \, . 
\label{eq:expESNM}
\end{equation}

The value of the incompressibility of infinite nuclear matter at 
saturation density has been the subject of considerable 
debate for several decades. 
It can be extracted either from measurement of energies of giant 
monopole resonances (GMR) in spherical nuclei or calculated 
theoretically in non-relativistic and relativistic models, 
typically involving mean-field plus RPA 
(see e.g. Refs.~\cite{blaizot1980,shlomo2006,stone2013}). 
The consensus has gravitated to a value of $K_{\rm 0}$ = 240 $\pm$ 20 MeV, 
as calculated in non-relativistic approaches, although somewhat higher 
values are predicted in relativistic models. Recent re-analysis of 
experimental data on GMR energies in nuclei with 56 $<$ A $<$ 208, 
in an empirical approach~\cite{stone2013} showed that $K_{\rm 0}$ 
critically depends on properties of the nuclear surface and the 
most likely values of $K_{\rm 0}$ are between 250 and 315 MeV. 

There is no rigorous constraint available for the skewness coefficient 
except for the results of Farine {\it et al.}~\cite{farine1997}. They 
obtained a model dependent value $K' = 700 \pm 500$ MeV from an 
analysis of the nuclear breathing mode, using a selection of Skyrme forces. 

We now give details of how the optical potentials of hyperons embedded in symmetric nuclear matter 
are calculated. We refrain from discussing these results until
Section \ref{subsec:gbem} as these potentials are intimately connected to the particle content of matter in
generalised beta equilibrium
and are therefore more naturally discussed there.
In many works optical potentials for the hyperons in symmetric nuclear matter
are evaluated and used to constrain hyperon coupling constants. 
In the QMC model these couplings are derived within the model.
We make the following approximation to evaluate the optical potentials in symmetric nuclear matter at saturation density.
For each hyperon a small number density is chosen, so that we can evaluate 
their chemical potential numerically via Eq.~(\ref{eq:chempots}). 
A small density means that this chemical potential is approximately the energy 
of a zero momentum hyperon embedded in symmetric nucleon only matter.
We can then calculate the optical potentials  by
$U_{i}(\rho_{0}) = \mu_{i} - M_{i}$.
These values are tabulated in Table~\ref{table:couplings} for the $\Lambda$, $\Sigma^{-}$ and $\Xi^{-}$ hyperons.

In PNM, $\rho_{\rm n}$ = $\rho$ and $\rho_{\rm p}$ = 0. Although PNM 
does not exist in nature, it is seen as a first approximation to matter 
in the outer core of neutron stars at densities higher 
than $\rho_{\rm 0}$. The density dependence of the energy per particle 
of PNM is poorly known, except for the fact that PNM does not bind 
\emdash i.e. the energy per particle is positive at all densities. 

At very low densities, below $\sim$ 0.1 $\rho_{\rm 0}$, experiments 
with cold Fermi atoms have yielded information about strongly 
interacting fluids, similar to low density matter in neutron star 
crusts. Dutra {\it et al.}~\cite{Dutra:2012mb} studied these constraints 
in detail. In this work we concentrate on the higher density region, 
above $\sim$ 0.1 $\rho_{\rm 0}$, as the QMC model may have limited 
applicability at very low densities.  In the absence of
      experimetal data in this density region we can only use theory
      for a comparison.

 Very recently, 
Tews {\it et al.}~\cite{Tews:2012fj} presented the first complete 
N$^{\rm 3}$LO calculation of the PNM energy, and 
Hebeler and Furnstahl~\cite{hebeler2013} investigated the energy 
per particle in PNM at sub-saturation densities using two- and 
three-nucleon CEFT interactions that were consistently evolved 
within the framework of the similarity renormalization group. 
We compare their results with the QMC predictions 
in Fig.~\ref{fig:PNMEnergyPerPart}. 
Clearly the QMC prediction for the density dependence of the energy 
per particle in PNM is very similar to that of 
Tews {\it et al.}~\cite{Tews:2012fj} at sub-saturation density, 
with a somewhat steeper increase at densities above saturation.

An interesting connection has been made between the pressure in 
the PNM neutron skin in heavy nuclei and the radius and crust 
thickness of a cold neutron star~\cite{horowitz2000}. 
Thus a microscopic theoretical calculation of the PNM pressure 
became of interest, in particular at sub-saturation densities. 
Tsang {\it et al.} (see Fig.~4 and related references in Ref.~\cite{Tsang:2012se}) 
collected several recent calculations of the PNM pressure as a function 
of particle number density. We show in Fig.~\ref{fig:PNMpressure} 
a selection of the models; Bruckner-Hartree-Fock (BHF) with Av18 
two-body potential~\cite{vidana2009}, Quantum Monte Carlo (QuMoCa) 
with Av8' two-body potential~\cite{gandolfi2012} and 
CEFT~\cite{hebeler2010b}. The main uncertainty in these calculations 
is the strength of three-body forces, which clearly make a significant 
contribution to the total pressure in these models (compare the left 
and right panels of Fig.~\ref{fig:PNMpressure}, with the QMC result  
shown in the right panel). The QMC model, which naturally includes 
three-body forces without additional parameters (see Sec.~\ref{sec:dis}), 
indicates a somewhat faster growth of pressure with increasing 
density than the other three-body interactions.

Limits for the pressure-density relationship in SNM and PNM in the 
density region \mbox{2~{--}~5~$\rho_{\rm 0}$} have been inferred from a 
comparison of experimental data on matter flow in energetic heavy 
ion collisions and predictions of a dynamical transport theory 
by Danielewicz {\it et al.} (see Ref.~\cite{danielewicz2002} and references 
therein). The matter created in the collision, 
lasting $\sim$ 10$^{\rm -23}$s at an incident 
kinetic energy per nucleon varying from about 0.15 to 10 GeV per 
nucleon, was modeled as consisting of stable and excited 
nucleons ($\Delta$ and N*) as well as pions. The basic constraints 
on this matter are charge symmetry and strangeness conservation 
(although in this case the strangeness is zero). This is in contrast 
to matter in cold neutron stars, constrained by charge neutrality 
and generalized beta-equilibrium, where strangeness 
will not be conserved. 

The transport theory was extrapolated to cold symmetric and pure 
neutron matter, with the latter augmented by empirical symmetry 
pressure~\cite{danielewicz2002}.  We show in Fig.~\ref{fig:SNMPNMEoS} 
the pressure versus density for SNM 
and PNM, 
as 
predicted in different scenarios the QMC model in this work. The standard QMC model 
is consistent with the suggested constraints but at the upper end 
of the range determined in \cite{danielewicz2002}.

\subsection{\label{subsec:anm} Asymmetric nuclear matter}
Our knowledge of asymmetric nuclear matter is rather limited, 
mainly because of a still inadequate understanding of the 
symmetry energy which describes the response of forces 
acting in a nuclear system with an excess of protons and neutrons. 
This is an important property of highly asymmetric systems, 
such as heavy nuclei and the nuclear matter found in neutron stars, 
and is defined as
\begin{equation}
\mathcal{S}(\rho) = 
\frac{1}{2}\frac{\partial^2 E}{\partial \beta^2}\big|_{\rho,\beta=0} \, , 
\label{eq:symE}
\end{equation}
where $\mathcal{S}(\rho)$ is equal to the asymmetry coefficient 
in the Bethe--Weisacker mass formula in the limit  
A $\rightarrow \infty$ \cite{Glendenning:2000wn}.

The definition of $\mathcal{S}(\rho)$ in Eq.~(\ref{eq:symE}) 
is related but not identical to the commonly 
used approximation as the difference
between the binding energy per baryon in PNM and SNM
\begin{equation}
S(\rho) = \mathcal{E}(\rho, \beta = 1)- \mathcal{E}(\rho, \beta = 0)   \ , 
\label{eq:difSym}
\end{equation} 
where the binding energy per baryon is 
\begin{equation}
\mathcal{E} = \frac{1}{\rho} \left( \epsilon_{\rm hadronic}-\sum_{B}M_{B}\rho_{B} \right) \ .
\end{equation}
This difference approximation is valid under two assumptions: 
(i) E($\rho, \beta$ = 0) is a minimum energy of the matter at 
a given density $\rho$ and thus in the expansion of E($\rho$, $\beta$) 
about this value with respect to $\beta$ the leading non-zero term 
is the second derivative term and 
(ii) all the other derivatives in the expansion are 
negligible~\cite{stone2007a}. In this work we consider 
Eq.~(\ref{eq:difSym}) only to examine the validity of this
approximation and to observe the impact of the Fock terms, 
specifically the tensor contribution, upon the symmetry energy. 

The density dependence of the symmetry energy can be expanded 
about its value at saturation $\mathcal{S}_{\rm_0}$ in terms 
of the slope $L_{0}$, curvature $K_{\rm sym}$ and skewness 
$Q_{\rm sym}$ (all evaluated at saturation density) as
\begin{equation}
\mathcal{S}=\mathcal{S}_{\rm_0} +L_{0}x+\frac{1}{2}K_{\rm sym}x^2+ 
\frac{1}{6}Q_{\rm sym}x^3+\mathcal{O}(x^4) \, , 
\label{eq:expansion}
\end{equation}
where 
\begin{eqnarray}
L(\rho)&=&3\rho\left(\frac{\partial \mathcal{S}}{\partial\rho}\right)\  ,  \  L_{0} \equiv L(\rho_{\rm o})\nonumber \  ,   \\
K_{\rm sym}&=&9\rho_{\rm o}^2\left(\frac{\partial^2 \mathcal{S}}{\partial\rho^2}\right)_{\rho=\rho_{\rm 0}} \   ,   \nonumber  \\
Q_{\rm sym}&=&27\rho_{\rm o}^3
\left(\frac{\partial^3 \mathcal{S}}{\partial\rho^3}\right)_{\rho=\rho_{\rm 0}}
\, .
\end{eqnarray}

We note that the curvature of the symmetry energy, $\mathcal{S}$, 
at saturation density in symmetric matter, is called here $K_{\rm sym}$, 
the symmetry incompressibility. It should not be confused with 
$K_\tau$, which is the isospin incompressibility, 
defined in ANM by Eq.~(\ref{ktau}).\\
The search for constraints on the symmetry energy and its slope, $L_{0}$, 
has received considerable attention during the last decade. 
Recently Tsang {\it et al.}~\cite{Tsang:2012se} evaluated constraints from 
a wide range of experiments. However, as again the symmetry energy 
is not measured directly but extracted from experimental data in a 
model dependent way, only limits on the symmetry energy can be 
established. One of the outcomes of the evaluation was a confirmation 
of a previously observed correlation between the value of 
$\mathcal{S}_{\rm 0}$ and its derivative $L_{0}$ at saturation density. 
Taking this correlation into account, the constraint centered 
on ($\mathcal{S}_{\rm 0}, L_{0}$) $\sim$ (32.5, 70) MeV, with the 
uncertainty in $\mathcal{S}_{\rm 0}$ allowing values 
30 $<$ $\mathcal{S}_{\rm 0}$ $<$ 35 MeV  and related values 
of $L_{0}$ in the range of $35 < L_{0} < 115$ MeV 
(see Fig.~2 in Ref.~\cite{Tsang:2012se} for more details). 

While theoretical predictions of $\mathcal{S}_{\rm 0}$ are also 
more or less confined to the range 30 to 35 MeV, the
calculated values of $L_{0}$, corresponding to the range of 
$\mathcal{S}_{\rm 0}$, vary widely. For example, the QuMoCa and 
CEFT models predict very similar low values of $L_{0}$, 
between $\sim$ 30 - 50 MeV~\cite{Tsang:2012se}. The best performing 
Skyrme forces, selected in  Ref.~\cite{Dutra:2012mb}, produce 
values of $L_{0}$ clustered around 50 MeV. On the other hand, relativistic 
mean field models show a much larger spread. The models which 
satisfied most of the constraints on the properties of nuclear matter, 
studied by Dutra {\it et al.}~\cite{Dutra:2013osa}, predicted $L_{0}$ in the 
range $\sim$ 50 -- 70 MeV. However, frequently used relativistic 
mean field model parameterizations, e.g. NL3, NL-SH, NLC, TM1 and TM2 
predict $L_{0}$ values of order $\sim$  110 -- 120
MeV~\cite{Sagawa:2007sp}.
Chen  {\it et al.} \cite{Chen:2009wv} found a linear the correlation
between K$_{sym}$ and $L_{0}$ for a specific selection of equations of
state. For a range of positive values of $L_{0}$ between about 30 -
120 MeV  K$_{sym}$ is between $\sim$ -200 and 100 MeV.

In the QMC model the isospin dependent part of the interaction
is mostly controlled by the exchange of the rho meson. For this
reason, here and in other works (e.g.~\cite{Glendenning:2000wn}) the
symmetry energy at saturation $\mathcal{S}_{\rm 0}$ = 32.5 MeV is 
used to fix the rho meson coupling
constant. The QMC result for $L_{0}$ is 84 MeV 
(see Table~\ref{table:couplings}), which is within the broader 
limits found by Tsang {\it et al.}~\cite{Tsang:2012se}, although outside 
their preferred range. 

We show the density dependence of the symmetry energy $\mathcal{S}$ 
and its slope, $L$, in Fig.~\ref{fig:SymSlopePlots} and the correlation 
between $\mathcal{S}_{\rm 0}$ and $L_{0}$ 
in Fig.~\ref{fig:LinearSL}. It can be seen that the linear 
relationship between $\mathcal{S}_{\rm 0}$ and $L_{0}$, observed in QuMoCa
calculations~\cite{Gandolfi:2011xu} and CEFT models~\cite{Tsang:2012se} 
is also predicted in this work, although at higher values of $L_{0}$ 
and a somewhat different incline. When the approximate expression is used to evaluate 
the symmetry
energy the linear relationship between $\mathcal{S}_{\rm 0}$ 
and $L_{0}$ is shifted
to values which are at most only a few MeV lower.

Another manifestation of isospin asymmetry in nuclear matter can be 
studied in Giant Monopole Resonance (GMR) experiments~\cite{stone2013}. 
The incompressibility of a finite nucleus  
is obtained, using sum-rule arguments, from the measured energy
$E_{\rm GMR}$ in spherical
nuclei~\cite{blaizot1980} as:
\begin{equation}
K(A,\beta)= M <R^2> E^{\rm 2}_{\rm GMR} \, .
\end{equation}
Here, M is the nucleon mass and $R$ is the 
rms \textit {matter} radius of the nucleus with mass number A. 
$K(A,\beta)$ can be expressed in a form of an expansion in terms 
of A$^{\rm -1/3}$ and $\beta$~\cite{blaizot1980}
\begin{eqnarray}
K(A,\beta)& = & K_{\rm vol} + K_{\rm surf}A^{-1/3} + K_{\rm curv}A^{-2/3}\\ \nonumber
          & + & K_{\tau}\beta^{\rm 2} + K_{\rm coul} 
\frac{Z^{\rm 2}}{A^{\rm 4/3}} + \cdots \,  , 
\label{ka}
\end{eqnarray}
where the symmetry related coefficient consists of the volume 
and surface components~\cite{blaizot1980,treiner1981,nayak1990}
\begin{equation}
K_\tau=K_{\tau,v}+K_{\tau,s}A^{-1/3} \,  , 
\label{ktau}
\end{equation}
with $K_{\rm \tau,v}$ ($K_{\tau,s}$ ) the volume (surface) symmetry 
incompressibility. 

The coefficient $K_{\tau,v}$ can be evaluated using 
\begin{equation}
K_{\tau,v}=\left(K_{\rm sym} - 6L_{0} -\frac{Q_{\rm 0}}{K_{\rm 0}}L_{0}\right) \, .
\end{equation}
Stone {\it et al.}~\cite{stone2013} analyzed all currently available 
GMR data in nuclei with 56 $<$ A $<$ 208 and found a 
limit -700 $\le K_{\tau,v} \le$ -372 {\rm  MeV}. 
The QMC result is K$_{\tau,v}$ = -431 MeV, which lies well 
within the experimental limits. 

\subsection{\label{subsec:gbem}Generalised Beta Equilibrium Matter and Neutron Stars}
In this section we study cold, asymmetric nuclear matter (ANM) which is 
expected to exist in the outer core of cold neutron stars. 

Dense matter just above the saturation density, when all nuclei are 
dissolved, forms a system of interacting nucleons and leptons. If this form 
of matter exists long enough on the time scale of weak interactions, 
$\tau \approx$ 10$^{\rm -10}$ s, beta-equilibrium (BEM) develops 
between beta decay $n \rightarrow p + e^{\rm -} + \tilde{\nu}$ and its 
inverse. When the density increases to about 2 -- 3 $\rho_{\rm 0}$ and 
because baryons obey the Pauli principle, it becomes energetically favorable for 
nucleons at the top the corresponding Fermi sea to convert to other baryons.
A generalized beta equilibrium (GBEM) develops with respect to all reactions involving either
weak or strong interactions, that lead to the lowest energy state.  
Only two quantities are conserved 
in GBEM - the total charge (zero in stars) and total baryon number. 
Strangeness is conserved only on the time scale of strong interaction, 
$\tau \approx$ 10$^{\rm -24}$s, and lepton number is conserved only on 
the time-scale of tens of seconds, because of the diffusion of 
neutrinos out of the star~\cite{Glendenning:2000wn}.  

To describe GBEM, it is convenient to use the chemical potentials of 
the participating particles. It can be shown that there are as many 
independent chemical potentials as the number of conserved quantities. 
Thus we need to choose just two, for example the chemical potentials of 
the neutron and electron.  Chemical potentials of all the other species 
in GBEM are then expressed via a relation
\begin{equation}
\mu_i = B_i \mu_n - Q_i \mu_e \, ,
\end{equation}
where the baryon number, $B$, is 0 or 1 and the charge number, $Q_{i}$, is 0 
or $\pm$1. Alternatively (and
equivalently), the chemical potentials can be related to Lagrange 
multipliers (as the degrees of freedom for charge
conservation ($\nu$) and baryon number conservation ($\lambda$)) in
order to solve the following system of equations 
\begin{eqnarray}
0 &=& \mu_{i} + B_{i}\lambda + \nu Q_i\ , \\
0 &=& \mu_{\ell}-\nu\ ,  \\
0 &=& \sum_{i} B_{i} \rho_{i}-\rho\ , \\
0 &=& \sum_{i} B_{i} \rho_{i}Q_{i} 
+ \sum_{\ell} \rho_{\ell}Q_{\ell}\  ,  \ 
\end{eqnarray}
to obtain the number densities
for each particle 
(\mbox{$i\in\{n,p,\Lambda,\Sigma^-,\Sigma^{0},\Sigma^+,\Xi^{-},\Xi^0\}$}
and \mbox{$\ell\in\{e^{-},\mu^{-}\}$}), $\rho_i$, as well as 
the Lagrange multipliers. At Hartree--Fock level, the following formulas to
numerically evaluate the chemical potentials, must be used to ensure we
encapsulate the Fock contribution to the energy densities correctly
\begin{equation}
\mu_{i} = \frac{\partial \epsilon_{\rm total}}{\partial \rho_{i}}\ , 
\quad \mu_{\ell} = \frac{\partial\epsilon_{\ell}}{\partial \rho_{\ell}} 
= \sqrt{k^{2}+m^{2}_{\ell}} \, .
\label{eq:chempots}
\end{equation} 

In Figs.~\ref{fig:BEMEoS} and \ref{fig:SpFr} 
we show the EoS (with various parameter variations) and the distribution of 
species in GBEM matter for the preferred scenario in this work.
We note that the pressure now involves the total energy
density (including leptonic contribution) $\epsilon_{\rm total} =
\epsilon_{\rm hadronic} + \epsilon_\ell$:
\begin{equation}
\label{eq:deftotpressure}
P_{\rm total} = \rho^{2}\frac{\partial}{\partial \rho}
\left(\frac{\epsilon_{\rm total}}{\rho}\right) 
= \sum_{i} \mu_{i}\rho_{i}-\epsilon_{\rm total} \, .
\end{equation}
The kinks in pressure in Fig.~\ref{fig:BEMEoS} appear at hyperon thresholds. 
A comparison between calculations for either Hartree alone, 
Hartree -- Fock with only the Dirac piece of the coupling to vector mesons, 
or the full model highlights the importance of the Fock terms at high density. 
As compared to the EoS of matter in which the hyperons 
are not included above their natural thresholds and nucleons are assumed 
to be the only baryons up to densities $\sim$ 5--6 $\rho_{\rm 0}$, 
the pressure in GBEM increases with density more slowly. 
 It is challenging to produce reasonable scenarios where the empirical constraints
 are met and the pressure still increases fast enough to support high mass, cold neutron 
stars, as will be discussed in the next section.

In Fig.~\ref{fig:SpFr} the particle content of GBEM matter and corresponding
Fock energy contributions is displayed for three scenarios, ``Standard", ``Eff. Proton Mass"
and ``$\Lambda=2.0$, $g_{\sigma Y}\times 1.9$".
In the ``Standard" scenario, the first hyperon to appear is $\Sigma^{-}$, at $0.46$fm$^{-3}$,
 followed by $\Xi^{-}$ at $0.47$fm$^{-3}$. The $\Sigma^{-}$ is quickly replaced by the $\Xi^{-}$,
  which is then followed by the appearance of $\Lambda$ at $0.74$fm$^{-3}$ and then $\Xi^{0}$ at $0.97$fm$^{-3}$. Since the latter is above the maximum density
reached in any of our realistic model variations it is largely irrelevant.
We show in Fig.~\ref{fig:ChemPots} 
that the $\Lambda$ chemical potential 
approaches and meets the neutron chemical potential, meaning that 
it is energetically favourable for it to appear. On the other
hand, for the $\Sigma^{-}$ we see that at low density it is more
favourable than the $\Xi^{-}$, while beyond $\sim 0.4$~fm$^{-3}$ this
is no longer so.

 Once the full Fock terms are included the results for the standard 
scenario are no longer consistent with the known values of the phenomenological hyperon optical potentials. 
This is because of
a change in the ratio of the scalar to vector coupling, effectively leaving the $\Lambda$ hyperon  unbound. 
The additional attraction generated by the Fock terms, especially the $\rho$ tensor contribution, 
has altered the coupling constants such that the $\omega$ coupling is larger. 
 This effect of an increase in the vector coupling is illustrated by the larger maximum neutron
 star masses, which also correspond to poor results for the hyperon optical potentials.
 
  In the work
of Miyatsu {\it{et al.}} the scalar couplings from the QMC model were not used. Instead they rescaled the
scalar coupling of each hyperon to obtain an acceptable optical potential. 
We consider the possibility of rescaling the scalar coupling reasonable, as the bag model used is a very
simple model of the baryons in which only the light quarks participate in the interaction.  An amplification
of only the hyperon scalar couplings of 30\% is considered in ``$\Lambda = (1.1,1.3) $, $g_{\sigma Y}\times 1.3$".  This improves the predictions of the optical potentials, binds the $\Lambda$ hyperon and maintains a repulsive potential for the $\Sigma^{-}$ hyperon.
In doing this the optical potentials are closer to the
values extracted from experimental studies of hypernuclei, but the EoS of $\beta$-equilibriated
matter is much softer, $\Lambda$ and $\Sigma^{-}$ both appear. 
In the scenario ``$\Lambda =2.0$, $g_{\sigma Y}\times 1.9$", we meet
both the constraints of phenomenological hyperon optical potentials and high mass neutron star observations. In this scenario we increase the form factor cutoff and hence the strength of the Fock terms forcing the vector coupling to become larger and then rescale the hyperon scalar coupling. In this very phenomenological scenario we obtain reasonable values for the optical potentials and
still obtain high mass neutron stars.

In the scenario ``Eff. Proton Mass" the ratio of tensor to vector coupling
is rescaled using the effective proton mass in Eq.~(\ref{eq:kapRescale}) as opposed to 
the free proton mass. This is a naive way to introduce a scalar dependence
into the Pauli term coupling. This substitution effectively increases the strength
of the Pauli term due to the reduction of the proton mass. The change in strength of the 
tensor coupling has a significant impact on the composition.  
It's  impact on the results for neutron stars will be discussed in the Sec. III F-E.
This particle content is different from our standard scenario and
 most other models, which generally find that either the $\Lambda$
or $\Sigma^{-}$ appears first. The increased strength of the tensor contribution, and hence attraction,
has increased the vector coupling and as a consequence the $\Lambda$ is not bound at saturation density 
in symmetric nuclear matter. This combined with the attraction from the Fock terms for the 
$\Xi$'s makes them more energetically favourable than $\Lambda$ or the $\Sigma^{-}$.

At the Hartree level the ``hyperfine'' interaction from one
gluon exchange makes the $\Lambda$ more energetically favourable than
the $\Sigma^{-}$, providing a source of attraction for the former and
repulsion for the latter. This has been shown at the Hartree level in
the QMC model to suppress the appearance of $\Sigma^{-}$ hyperons in
GBEM matter~\cite{Carroll:2010ex}. This can
also be considered a qualitative explanation for the absence of medium
to heavy $\Sigma$ hypernuclei~\cite{Guichon:2008zz}. 

We consider the extreme scenario ``$\Lambda=2.0$, $g_{\sigma Y}\times 1.9$ 
where we maintain a large vector coupling by increasing the form factor cutoff, effectively 
increasing the strength of the Fock terms and we rescale the scalar coupling to obtain reasonable
values for the phenomenological  hyperon optical potentials, so that the $\Lambda$ is bound at
saturation density in symmetric nuclear matter.
Even though the $\Lambda$ feels a significant attraction at saturation density, it appears that it cannot compete with the attraction generated by the Fock terms at high density \emdash specifically the
tensor part \emdash for the $\Xi$. The contributions of the Fock energies is more significant
and the composition is similar to ``Eff. Proton Mass".

\subsection{\label{subsec:cns}Cold neutron stars}
In order to calculate neutron star properties, such as the total
gravitational mass, $M(R)$, and the baryon 
number, $A(R)$, within the stellar radius $R$, 
we solve the TOV equations~\cite{tov} for hydrostatic 
equilibrium of spherically symmetric (non-rotating) matter. 
Using the EoS calculated here, this is self-supported against
gravitational collapse. 
\begin{eqnarray}
\label{eq:TOV1}
M(R) &=& \int_0^R 4\pi r^2 \epsilon_{\rm total}\, dr\ , \\[2mm]
\label{eq:TOV2}
\frac{dP_{\rm total}}{dr} &=& -(\epsilon_{\rm total} 
+ P_{\rm total}) \frac{(M(r) + 4\pi r^{3}P_{\rm total})}{r^{2}(1 - 2M(r)/r)}\ , 
\\[2mm]
\label{eq:TOV3}
\frac{dA}{dr} &=& \frac{4\pi r^{2}\rho}{\sqrt{1-2M(r)/r}} \, .
\end{eqnarray}
In Eqs.~(\ref{eq:TOV1}--\ref{eq:TOV3}) we use units in which
\mbox{$G=1$}. The difference between the total gravitational mass and
baryonic mass within a radius $R$ is defined by $M(R) - A(R)M_N$. 

The EoS of GBEM is not valid in the outer regions (crust) of the star, 
where nuclei and nuclear processes become dominant. Following 
the customary procedure, we introduce 
a smooth transition between our EoS in  GBEM  
and the standard low density EoS of Baym, Pethick and 
Sutherland (BPS)~\cite{Baym:1971pw} at 
low density.

The relationship between stellar mass and radius, obtained as the 
solution of the TOV equations, Eqs.~(\ref{eq:TOV1}--\ref{eq:TOV3}),
is summarized in Table~\ref{table:couplings} and depicted in
Fig.~\ref{fig:MVsR}. We find that the predicted maximum masses for several of the scenarios, 
lie very close to the constraints set by  Demorest {\it et al.}~\cite{Demorest:2010bx}
of a (1.97 $\pm$ 0.04) M$_{\odot}$ pulsar, as well as 
the new constraint set by 
PSR J$0348+0432$ with a mass 
of 2.03 $\pm$ 0.03 M$_{\odot}$~\cite{Antoniadis:2013pzd}. 
The corresponding radii are
somewhat larger than that extracted from recent observations of 
Type I X-ray bursters (see e.g. Refs.~\cite{steiner2010,guillot2013}). 
Extraction of radii from observation is rather complicated and there are 
still many questions to be addressed. For example, 
Steiner {\it et al.}~\cite{steiner2010} analyzed observations of six 
low mass X-ray binaries (emitting X-rays regularly) and 
their statistical analysis yielded $R$ in the range 10--12 km for masses 
around $1.6 M_{\odot}$. However, the uncertainty in the relation between the 
extracted photospheric radius and the actual radius of the star remains  
large. The results of Guillot {\it et al.}, 
namely $R = 9.1^{\rm +1.3}_{\rm -1.5}$ km  (90\%-confidence), are based 
on observations of five quiescent low mass X-ray binaries (which emit 
X-rays only occasionally) under the assumption that the radius is 
constant for a wide range of masses.

Whilst the observations Refs.~\cite{Demorest:2010bx,Antoniadis:2013pzd} 
provide constraints on high mass neutron stars, 
the observation of the double pulsar J0737-3039 and its 
interpretation~\cite{podsiadlowski2005} offers a constraint on the 
neutron star EoS in a region of central densities $\sim$ 2 -- 3 $\rho_{\rm 0}$. 
The constraint concerns the ratio between the gravitational and 
baryonic mass of the star. The gravitational mass of pulsar B is 
measured very precisely to be $M_{\rm g} = 1.249 \pm 0.001 M_{\odot}$ and 
the baryonic mass depends on the mode of its creation, which can be modeled. 
If pulsar B was formed from a white dwarf with an O-Ne-Mg core in an 
electron capture supernova, with no or negligible loss of baryonic mass 
during the collapse, the newly born pulsar should have the same baryonic mass 
as the progenitor star. Podsiadlowski {\it et al.}~\cite{podsiadlowski2005} 
estimated the baryonic mass of the pulsar B to be between 1.366 
and 1.375 $M_{\odot}$. Another simulation of the same process,
by Kitaura {\it et al.}~\cite{kitaura2006}, gave a value for the baryonic 
mass of 1.360 $\pm$ 0.002 $M_{\odot}$. We show in Fig.~\ref{fig:MgVsMb} 
the QMC result, which supports the model of Kitaura {\it et al.}, accepting 
some small loss of baryonic mass during the birth of pulsar B.\\

\subsection{\label{subsec:sen}Sensitivity to parameter variation}

Our calculations for the Hartree--Fock QMC model follow similar
lines to Refs.~\cite{RikovskaStone:2006ta,Guichon:2006er,Krein:1998vc}
in that in each case an approximation is made for the Fock terms. More
specifically, in our calculation of the Fock terms we omit energy
transfer in the meson propagator (meson retardation effects). We also omit the
modification of momenta because of the vector component of the self
energy, which has been shown to be small in Refs.~\cite{Krein:1998vc} and
\cite{Miyatsu:2011bc}. We include the tensor interaction
in the Fock terms, with a common form factor, which has a dipole form.
The lowest mass, $\Lambda$, for that cut-off, 
which should be larger than the masses 
of the mesons included, is 0.9 GeV. This is taken as our standard or baseline scenario value  for 2 reasons:\\

 (i)  The incompressibility $K_{0}$ rises as $\Lambda$ is increased.
In the range $\Lambda = 0.9$ -- $2.0$ GeV, for the scenarios considered $K_{0}$ 
remains within the 
 range $250\leq K_{0}\leq 315$MeV, which was the constraint derived in \cite{stone2013}.
\\

(ii) Increasing the form factor cutoff $\Lambda$, effectively increases the strength of
the Fock terms, for which the $\omega$ and $\rho$ mesons contribute a significant
attraction once contact subtraction has been performed. To obtain the saturation 
properties of SNM, one must compensate for  this additional attraction, resulting in
a larger vector coupling. If the vector couplings of the hyperons are simply related to
to the vector couplings of the nucleons by Eq. (\ref{eq:vectcc}), the results for the hyperon optical potentials at 
saturation density in SNM are not consistent with the values extracted from hypernuclear 
experiments largely because of the change in  the ratio of the scalar to vector couplings. \\

We demonstrate the effect of changing the value of $\Lambda$
between 0.9 -- 1.3 GeV in the subsequent scenarios (lines 2 - 5) in Table~\ref{table:couplings} which differ
from the standard one only by the value of $\Lambda$. We observe a
minor increase in $K_{0}$ and $L_{0}$ which both remain within the
empirical expected range and an increase in the maximum
mass the the neutron star by $\sim$ 8\%.

However, once the full Fock terms are included, the results for the standard 
scenario, even with variable $\Lambda$, are not consistent with values
of the phenomenological hyperon optical potentials extracted from experiments. This
is because of a change in the ratio of the scalar to vector coupling,
leaving effectively the $\Lambda$ hyperon unbound. The additional attraction generated by the Fock terms, especially the $\rho$ tensor contribution, 
has altered the coupling constants such that the $\omega$ coupling is
larger.

In the extreme scenario ``$\Lambda=2.0$, $g_{\sigma Y}\times 1.9$" discussed in Sec.~\ref{subsec:gbem}, we meet
both the constraints of phenomenological hyperon optical potentials and high mass neutron star observations.

In the scenarios ``Eff. Proton Mass", ``Eff. Proton Mass, $\Lambda$ =
1.1" and ``Eff. Proton Mass + $\delta\sigma$'' (lines 11--13 of Table \ref{table:couplings})  the ratio of tensor to vector coupling
is rescaled using the effective proton mass in Eq.~(\ref{eq:kapRescale}) as opposed to 
the free proton mass. This is a simplified way to introduce a scalar dependence
into the Pauli term coupling. This substitution effectively increases the strength
of the Pauli term due to the reduction of the proton mass. The change in strength of the 
tensor coupling has a significant impact on the composition.  It causes a significant
increase in $K_{0}$ as $\Lambda$ takes on larger values. 
Indeed, as we see in Table~\ref{table:couplings}, 
$K_0$ rises above 311 MeV for $\Lambda$ 
greater than 1.1 GeV.  Similar observations apply for the slope of the symmetry energy
at saturation density, $L_{0}$. Because of the increased vector
coupling, the maximum mass of the neutron star
is significantly increased, but the hyperon optical potentials remain
at variance with expected values.

The contribution to the mean scalar field arising from the Fock terms
is incorporated in the cases denoted ``Fock $\delta \bar{\sigma}$''
and ``Eff. Proton Mass $+\ \delta\sigma$". 
When applied to neutron star properties it negligibly increases the maximum mass
in our baseline scenario and increases it  by a few percent when a
scalar dependence is introduced into the Pauli term, to just below
2$M_\odot$. 

The tensor couplings used in this work, arising from the underlying MIT bag 
model, are consistent with Vector Meson Dominance (VDM) and 
hence our tensor couplings are calculated from the experimental 
magnetic moments. Purely as a test of the effect of a variation in 
those couplings we arbitrarily 
 took the ratios of tensor to vector couplings of all baryons from 
 the Nijmegen potentials (Table VII of Ref. \cite{Rijken:2010zzb}), 
 where there is a larger value of $f_{\rho N}/g_{\rho N} = 5.7$. 
These were 
also used by Miyatsu {\it et al.}~\cite{Miyatsu:2012xh,Miyatsu:2011bc}. 
This variation, denoted ``Increased $f_{\rho N}/g_{\rho N}$'', produced 
an EoS for GBEM which was indistinguishable from our standard result

In scenarios ``Dirac only, Hartree only and Nucleon only' we show
results of the QMC calculation with the same parameters as the
standard set but leaving out the Pauli part of the Fock term, the full
Fock term and the hyperons, respectively. These results are
particularly useful for understanding of the role of individual terms
in the QMC Lagrangian.

The last four scenarios in Table~\ref{table:couplings} document the effect of changes in the value of
the free nucleon radius and the evaluation of the symmetry energy
from the difference formula Eq. (\ref{eq:difSym}) ``App.' and from the
second derivative the the energy per particle ``S$_{\rm 0}$ = 30.0'.
Neither effect changes significantly the properties of GBEM matter and
neutron stars.\\

\subsection{\label{subsec:comp}Comparison with other models}

The Hartree -- Fock calculation in Ref.~\cite{Massot:2012pf} 
differs considerably from that presented here, as well as from that in
Refs.~\cite{Katayama:2012ge,Miyatsu:2012xh,Miyatsu:2011bc,Miyatsu:2010zz}. 
The first and major difference is that the tensor interaction of the
baryons is ignored, whereas in 
Refs.~\cite{Katayama:2012ge,Miyatsu:2012xh,Miyatsu:2011bc,Miyatsu:2010zz}
and in our work it is found to have a very significant effect.
A second difference between Ref.~\cite{Massot:2012pf}, our work, and
Refs.~\cite{Katayama:2012ge,Miyatsu:2012xh,Miyatsu:2011bc,Miyatsu:2010zz} 
is that in their preferred QMC scenario (QMC-HF3) 
they artificially adjust a parameter, 
$C$, which is related to the scalar polarisability, 
to obtain a lower value for the incompressibilty. 
This represents a dramatic change in the model.

The masses of the baryons in the QMC
model are determined by the bag equations and the scalar coupling
is calculated directly from the density dependence of the baryon mass in-medium.
Thus, changing $C$, or
equivalently the scalar polarisability, changes the mass and the
density dependent coupling in a manner which is inconsistent 
with the traditional form of the QMC model~\cite{Saito:2005rv}. In this manner the many
body interaction is also being changed through the density dependent scalar
coupling. Their QMC-HF3 variation gives an incompressibility of
$K=285$~MeV and a very low prediction for the maximum mass of neutron
stars, $M = 1.66~M_{\odot}$. In our Dirac-only variation we find a
slightly larger value for the incompressibility, $K=294$~MeV with a
maximum stellar mass of $M =1.79~M_{\odot}$. Other variations were
considered in Ref.~\cite{Massot:2012pf} where they do not modify $C$:
one where they calculate fully relativistic Fock terms, and another
where they make a non-relativistic approximation to the Fock
terms. These variations both produce maximum masses of neutron stars
of $M=1.97~M_{\odot}$.

Refs.~\cite{Katayama:2012ge,Miyatsu:2012xh,Miyatsu:2011bc,Miyatsu:2010zz}
carry out a relativisitic
calculation in which they treat the Fock and Hartree terms on the
same level. More precisely they calculate 
self-energy contributions arising from both terms and these self energies modify
the baryon mass, momentum and energy. They include the tensor
interaction, subtract contact terms, and consider two variations of
the bag model. In their first paper  \cite{Miyatsu:2011bc}  they used much larger  
values for the tensor couplings without form factors. In the later
paper  \cite{Katayama:2012ge} they include the effect of form factors, ignoring effects of
meson retardation (as we do) but with a lower cutoff mass, i.e.  $\Lambda =
0.84~{\rm GeV}$. The latter had the effect of keeping the 
incompressibility from being too large. 
Their conclusions are very similar to our own, in that they find that
the tensor terms provide a source of attraction and that overall the Fock
terms enhance the maximum neutron star mass.

The maximum stellar masses in their first paper \cite{Miyatsu:2011bc}  are
 larger than those in their second paper \cite{Katayama:2012ge}, almost
certainly because the inclusion of the form factor decreases the
effect of the Fock term at high density. They consider two variations
of the QMC model: one with, and one without the pion contribution in the bag
(CQMC) which tends to give a slightly stiffer EoS, because of its
effect on the baryon masses. For QMC they obtain $M=1.86~M_{\odot}$,
$R=11.2$~km, and for CQMC $M=1.93~M_{\odot}$, $R=11.5$~km for the
maximum stellar mass solutions. 
Despite the differences in how we
handle the Fock terms and their use of larger tensor couplings and 
more phenomenological hyperon couplings, we are led
to the same conclusions about the importance of the tensor
contribution. We also find a very similar particle
content in scenarios where the Fock terms are quite strong, such
as the ``Eff. Proton Mass" and ``$\Lambda=2.0$, $g_{\sigma Y}\times 1.9$" scenarios, where the $\Xi^{-}$ is the first hyperon to appear. 

\section{\label{sec:dis}Discussion}
In order to treat the equation of state of matter at the densities 
typical of neutron stars one must treat the motion of the 
baryons relativistically. The quark-meson coupling (QMC) model not 
only does that but it self-consistently treats the in-medium 
changes in baryon structure induced by the large scalar mean fields 
generated in such matter. As we have explained, those changes, 
which may be represented by the corresponding scalar polarisabilities, 
lead naturally to predictions for the three-body forces between not 
just the nucleons but the nucleons and hyperons as well as hyperons, 
without additional parameters. This widely used approach has been 
extended here to include the effect of Fock terms arising from 
the tensor (or Pauli) couplings of the baryons to vector mesons, 
especially the $\rho$.

The results for a comprehensive set of nuclear matter properties, 
including $K_0$, $L_{0}$, $K_{sym}$, $Q_0$ and $K_{\tau, v}$ have been 
studied in detail. 
The model prediction for the incompressibility lies within the range extracted
from experimental data for most model variations considered.
While the incompressibility is increased 
by this addition in some cases and tends to lie at the mid to top end of the 
acceptable range, it serves as a useful constraint on the 
additional mass parameter, $\Lambda$, associated with the form factor  
that appears at the meson-baryon vertices (the latter only being needed  
when the Fock terms are computed).
The modest variation of the nuclear matter observables with
this parameter (which must lie above the masses of the exchanged mesons 
included in the theory) is illustrated in Table~\ref{table:couplings}.
Increasing $\Lambda$ beyond $0.9$~GeV raises the incompressibility and in
the case denoted ``\mbox{Eff. Proton Mass, $\Lambda =1.1~{\rm GeV}$}" it is close to the
limit \mbox{$K_{0} < 315$~MeV}. 

The symmetry energy and its slope are noticeably influenced by the Fock terms, 
specifically curvature is introduced into these quantity through the tensor interaction,
as can be seen in Fig.~\ref{fig:SymSlopePlots}.
 At saturation density we find in all cases that the
isospin incompressibility is within accepted constraint limits and 
while the slope of the symmetry energy is on the larger side,  
it does lie within the broad limits reported 
by Tsang {\it et al.}~\cite{Tsang:2012se}.

It is interesting to note that there is a satisfying level 
of consistency between theoretical predictions of N$^{3}$LO chiral
effective field theory and the QMC model results studied here for
densities of PNM up to and around nuclear matter density.
Above saturation density a slightly higher energy per
particle as a function of density is found here. It is
also found that the natural incorporation of many body forces in the QMC
model tends to produce a somewhat stiffer PNM EoS above saturation density 
 than other models including 3-body forces.

Even at densities above three times nuclear matter density, 
the nucleon Fock terms are found to contribute significantly to 
the EoS and the corresponding 
attraction is what is responsible for the increased pressure and
larger maximum stellar masses in several scenarios. This can be seen in
{}Fig.~\ref{fig:MVsR}, where there is a clear transition from a Hartree
QMC calculation to a Hartree--Fock calculation with no tensor
interaction (Dirac-only; no Pauli term), to our ``Eff. Proton Mass" calculation
(Dirac and Pauli (with scalar dependence) terms). In these three variations, and those with
increasing form factor mass, $\Lambda$, the maximum stellar 
mass increases because of 
the increased vector coupling and pressure coming from the Fock terms. This increased
pressure arises mainly from the $\rho$ meson contribution. 
As we can readily see in Table~\ref{table:couplings} and Fig.~\ref{fig:MVsR}, the value of
$\Lambda$ cannot be varied far in the ``Eff. Proton Mass" calculations. Indeed, in that case, the 
incompressibility is already as high as it can be.
The maximum neutron star mass, for our ``Standard" scenario is approximately the
same as the ``Dirac Only" scenario because of the change in composition, where in the latter
the appearance of $\Sigma^{-}$ is avoided and only the $\Lambda$ and $\Xi^{-}$ followed by the $\Xi^{0}$ appear. 
Even with the brief appearance of an additional hyperon in our baseline scenario, the
value of $M_{\rm max}$ is 
still slightly larger because of the tensor interaction. 
We see that the maximum neutron star mass, for 
the case of nuclear matter in beta-equilibrium where hyperons must appear, 
lies in the range 1.80 to 2.07$M_{\odot}$.

The EoS and the maximum masses of the corresponding neutron stars 
are insensitive to the choice of the larger $\rho$ tensor couplings 
used, for example,
by Miyatsu {\it et al.}~\cite{Miyatsu:2011bc}. 
Similarly, modest variations in the radius of the free nucleon 
have only very minor effects on these quantities.
Finally, we note that the correction ($\delta \bar{\sigma}$) 
to the scalar mean field
arising from the Fock terms has a negligible effect on the incompressibility
in our baseline scenario. On inclusion of a naive scalar dependence into
the Pauli term it decreases the incompressibility by
$12$~MeV, yet other observables remain largely unaltered by this
addition.

This, plus the dependence of the incompressibility and 
maximum mass on $\Lambda$
 , leads us to conclude that the Hartree-Fock model used here with only $\sigma$, $\omega$, $\rho$ and $\pi$ mesons  can only reproduce nuclear matter properties, phenomenological hypernuclear optical potentials and massive neutron star observations if there is significant rescaling of the hyperon coupling constants. Allowing for the rescaling of hyperon couplings we conclude that the maximum mass allowed in the model lies in the range $1.8-2.1M_{\odot}$.

It is the treatment of the lightest mesons that is the most important, and the inclusion of heavier mesons would necessarily be more model dependent.  For this reason, in this work we have restricted ourselves to just $\sigma$, $\omega$, $\rho$ and $\pi$ mesons.
The model could be extended to include mesons containing strange quarks, of which the next lightest mesons are $K(495)$ and $K^{\ast}(895)$.
These mesons will induce mixing in the baryon octet, possibly changing the composition of matter in 
generalized beta-equilibrium. These mesons will be studied in a future work. Heavier mesons such
as the hidden strangeness vector meson $\phi(1020)$ have been considered in other works Ref.\cite{Weissenborn:2011kb,Weissenborn:2011ut} which
have found that they can produce extra vector repulsion delaying the onset of hyperons.
It should be noted that with every new meson that is included more parameters must be introduced into the model. 

For the matter considered in the present paper we take the view that hadrons remain the relevant degrees of freedom. Transitions to quark matter have been studied by many authors, see Refs.\cite{Masuda:2012ed,Masuda:2012kf,Blaschke:2013rma} for recent accounts. Such a transition may indeed be possible in the interior of neutron stars. We will investigate such a transition in a future work.

We stress that the QMC model does not predict a significant abundance of
$\Sigma$ hyperons at any density where the model can be considered
realistic and they are completely absent in model variations compatible with large neutron
star mass observations. This is in contrast to a number of other relativistic
models which do predict the $\Sigma$ threshold to occur, even prior to
that of the $\Lambda$~\cite{SchaffnerBielich:2010am,Weber:2005}. We
note that Schaffner-Bielich~\cite{SchaffnerBielich:2010am} considered
a phenomenological modification of the $\Sigma$ potential with
additional repulsion, which significantly raised its threshold
density. In the case of the QMC model the physical explanation of the
absence of $\Sigma$-hyperons is very natural, with the mean scalar
field enhancing the repulsive hyperfine force for the in-medium 
$\Sigma$ (recall that the hyperfine splitting, which arises from 
one-gluon-exchange, determines the free $\Sigma$--$\Lambda$ mass
splitting in the MIT bag model).

Purely for comparison purposes, we also include a nucleon-only scenario, in
which hyperons are artificially excluded. In this case the EoS is
increasingly stiffer at densities above $0.4~{\rm fm}^{-3}$, leading
to a large maximum stellar mass of $2.10~M_{\odot}$, consistent with
many other nucleon-only models.

It is worth remarking that upon inclusion of the tensor coupling, the
proton fraction increases more rapidly as a function of total baryon
density. This is likely to increase the probability of the direct
URCA cooling process in proto-neutron stars. As a further
consequence, the maximum electron chemical potential is increased in
this case, which may well influence the production of $\pi^{-}$ and
$\bar{K}$ condensates. Changes to the $\Lambda$ threshold (it occurs at
higher density with lower maximum species fraction) reduce the
possibility of H-dibaryon production as constrained by
beta-equilibrium of the chemical potentials.

In summary, taking into account the full tensor structure of the
vector-meson-baryon couplings in a Hartree--Fock treatment of the QMC
model results in increased pressure at high density -- largely
because of the $\rho N$ tensor coupling -- while maintaining
reasonable values of the incompressibility at saturation density. The
conceptual separation between the incompressibility at saturation
density and the slope of the symmetry energy or `stiffness' at higher
densities is critical. It is the latter that leads to neutron stars
with maximum masses ranging from $1.8~M_{\odot}$ to $2.1~M_{\odot}$,
even when allowance is made for the appearance of hyperons. This
suggests that hyperons are very likely to play a vital role as
consituents of neutron stars with central densities above three times 
nuclear matter density.

%
%
\section*{Acknowledgements}
%
%

JRS is pleased to acknowledge the hospitality of the CSSM at the
University of Adelaide, where this work was carried out. 
We thank Kai Hebeler for providing numerical data for the 
CEFT results in Fig.~\ref{fig:PNMEnergyPerPart}.
This work was supported by the University of Adelaide and the Australian
Research Council through grant FL0992247 (AWT) and through the ARC
Centre of Excellence for Particle Physics at the Terascale. 
DLW was supported by an ARC post graduate scholarship.
The work of KT was supported by the Brazilian Ministry of Science,
Technology and Innovation (MCTI-Brazil), and
Conselho Nacional de Desenvolvimento Cient{\'i}fico e Tecnol\'ogico
(CNPq), project 550026/2011-8.
\clearpage
\vspace{-5mm}
\section*{Appendix}
The integrands take the following form for $B=B'$
\begin{eqnarray}
\mathbf{\Xi}^{\sigma}_{B} 
& = & \frac{1}{2}\frac{\left(  g_{\sigma B}C_{B}(\bar{\sigma})F^{\sigma}(\mathbf{k}^{2})\right) ^{2}}{ E^{\ast}(\mathbf{p}')E^{\ast}(\mathbf{p})}\left\lbrace M^{\ast 2}_{B} + E^{\ast}(\mathbf{p}')E^{\ast}(\mathbf{p})-\mathbf{p'} \cdot \mathbf{p} \right\rbrace \Delta_{\sigma}(\mathbf{k}).\nonumber \\
\end{eqnarray}
Here for the vector meson integrands we denote $\eta =\omega,\rho$
\begin{eqnarray}
\mathbf{\Xi}^{\eta V}_{B} 
& = & -\frac{\left(  g_{\eta B}F^{\eta}_{1}(\mathbf{k}^{2})\right) ^{2}}{ E^{\ast}(\mathbf{p}')E^{\ast}(\mathbf{p})}  \left\lbrace  2M^{\ast 2}_{B} - E^{\ast}(\mathbf{p}')E^{\ast}(\mathbf{p}) + \mathbf{p}'\cdot \mathbf{p}  \right\rbrace\Delta_{\eta}(\mathbf{k}) \  ,  \\
\mathbf{\Xi}^{\eta VT}_{B} 
 & = &  \left(  g_{\eta B}\right) ^{2} \kappa_{\eta B} F^{\eta}_{1}(\mathbf{k}^{2}) F^{\eta}_{2}(\mathbf{k}^{2})  .\left\lbrace 
\frac{-3M^{\ast 2}_{B} +3E^{\ast}(\mathbf{p}')E^{\ast}(\mathbf{p})) -3\mathbf{p'}\cdot\mathbf{p}}{E^{\ast}(\mathbf{p}')E^{\ast}(\mathbf{p})} \right\rbrace\Delta_{\eta}(\mathbf{k}) \   ,  \\
\mathbf{\Xi}^{\eta T}_{B}
& = & - \frac{\left(  g_{\eta B}\kappa_{\eta B}F^{\eta}_{2}(\mathbf{k}^{2}) \right) ^{2}}{E^{\ast}(\mathbf{p}')E^{\ast}(\mathbf{p})}         
. \nonumber\\
&& .\left\lbrace \frac{(5 M^{\ast 2}_{B} -E^{\ast}(\mathbf{p}')E^{\ast}(\mathbf{p})+\mathbf{p'}\cdot\mathbf{p})}{4M^{\ast 2}_{B}} .(M^{\ast 2}_{B} -E^{\ast}(\mathbf{p}')E^{\ast}(\mathbf{p})+\mathbf{p'}\cdot\mathbf{p})\right\rbrace \Delta_{\eta}(\mathbf{k})
\end{eqnarray}
and for the pion
\begin{eqnarray}
\mathbf{\Xi}^{\pi}_{B}
 & = & -\frac{2M_{B}^{\ast 2}(\frac{g_{A}}{2 f_{\pi}}F_{\pi}(\mathbf{k}^{2}))^{2}}{ E^{\ast}(\mathbf{p})E^{\ast}(\mathbf{p}')} \left\lbrace M^{\ast 2}_{B}-E^{\ast}(\mathbf{p})E^{\ast}(\mathbf{p}')+\mathbf{p}'\cdot\mathbf{p}\right\rbrace  \Delta_{\pi} (\mathbf{k}).
\end{eqnarray}
where $E^{\ast}(\vec{p}) = \sqrt{\vec{p}^{\ 2}+M^{\ast  2}_{B}}$.
In the above integrands we expand the terms in the braces multiplied by the propagator
 to isolate the momentum independent pieces and multiply these contact terms by the variable $\xi$ which we 
use to investigate the consequences of contact subtraction. We emphasize here the importance of subtraction
of the momentum independent piece, which when transformed to configuration space corresponds to a delta
function.
 In this manner our subtraction is implemented by the variable $\xi$,  such that $\delta (\vec{r})
\mapsto \xi \times \delta (\vec{r} )$. The removal of the contact terms is a common procedure due to the fact that
these contact terms represent very short range, effectively zero range correlations between the
baryons, which is not consistent in this model which treats the baryons as clusters of quarks and not as point-like
objects. We give this explicitly for the Vector-Vector piece
of the vector mesons
\begin{eqnarray}
\frac{2M^{\ast 2}_{B} - E^{\ast}(p')E^{\ast}(p) + \vec{p'}\cdot \vec{p}}{\vec{k^{2}}+m^{2}_{\eta}} 
& =& \frac{2M^{\ast 2}_{B} - p'\cdot p}{\vec{k^{2}}+m^{2}_{\eta}} \nonumber\\  
& \simeq &  \frac{M^{\ast 2}_{B} - \frac{\vec{k}^{2}}{2}}{\vec{k^{2}}+m^{2}_{\eta}} \nonumber\\  
& = & \frac{M^{\ast 2}_{B}}{\vec{k}^{2}+m^{2}_{\eta}}-\frac{1}{2}\frac{\vec{k}^{2}}{\vec{k}^{2}+m^{2}_{\eta}}\nonumber\\
& = & \frac{M^{\ast 2}_{B} + m^{2}_{\eta}/2}{\vec{k}^{2}+m^{2}_{\eta}} -\frac{1}{2}\xi
\end{eqnarray}
the remaining subtractions follow in the same manner.

\clearpage
\begin{figure}
\includegraphics[width=\textwidth]{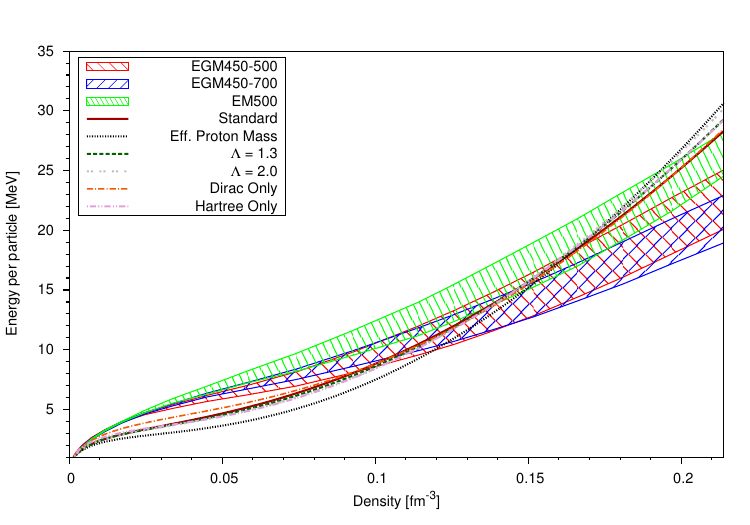} 
\caption{(Color online) Pure neutron matter energy per 
particle as a function of density as obtained in the present work,
in comparison with complete CEFT at N$^{\rm 3}$LO order -- for more 
details of the latter, see Ref.~\cite{Tews:2012fj}.}
\label{fig:PNMEnergyPerPart}
\end{figure}

\clearpage
\begin{figure}
\includegraphics[width=\textwidth]{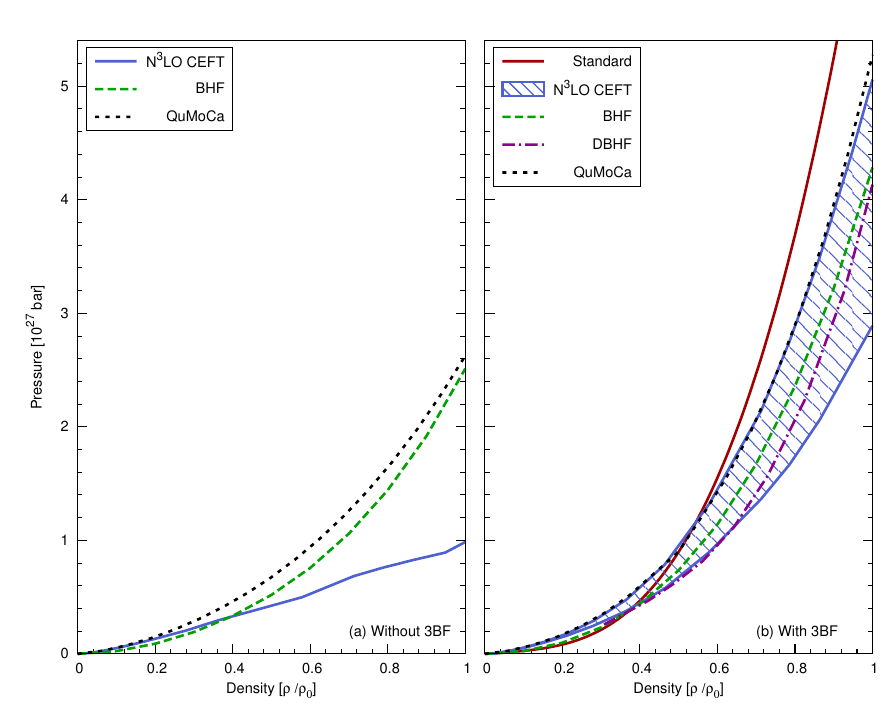} 
\caption{(Color online) Density dependence of pressure in PNM as predicted 
in BHF, DBHF, QuMoCa and CEFT with and without three-body forces.  
(a) Without three-body forces. (b) With three-body forces. The QMC model prediction is shown in (b). For more details see the text and ref.~\cite{Tsang:2012se}.}
\label{fig:PNMpressure}
\end{figure}

\clearpage
\begin{figure}
\includegraphics[width=0.8\textwidth]{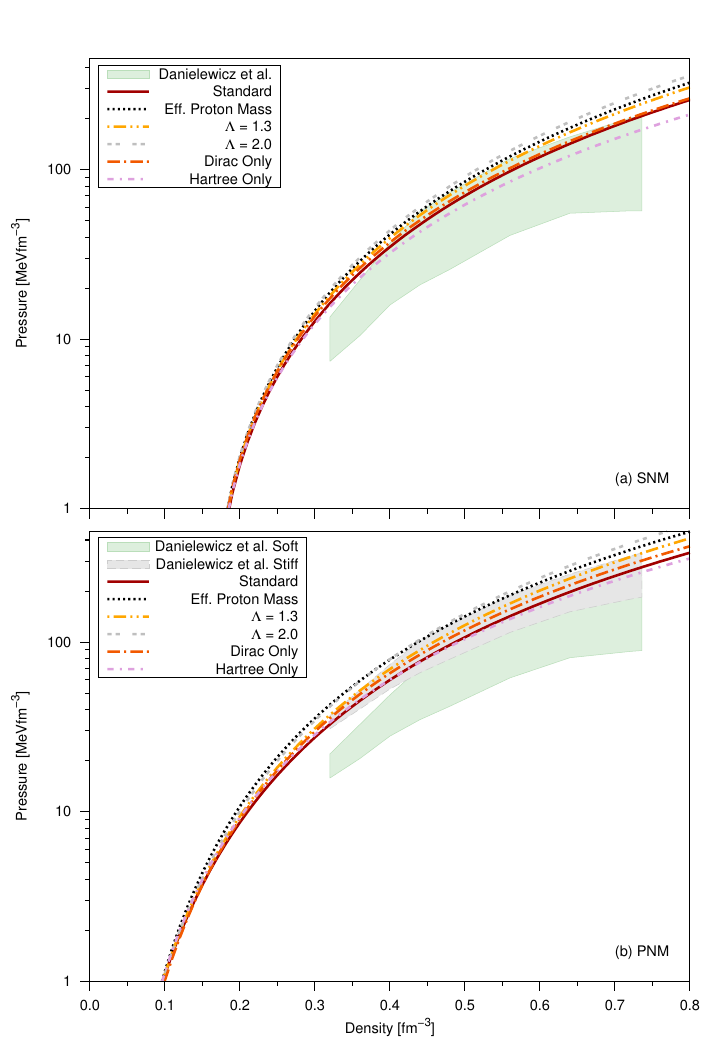} 
\caption{(Color online) (a) Pressure in SNM as a function of density 
as predicted in QMC model. The shaded area is taken from 
Ref.~\cite{danielewicz2002}.(b)  Pressure in PNM as a function of 
density as predicted in the QMC model. The upper and lower shaded 
areas correspond to two different estimates of the contribution of 
the symmetry pressure to the total pressure. 
{}For more detail see Ref.~\cite{danielewicz2002} }
\label{fig:SNMPNMEoS}
\end{figure}
\clearpage
\begin{figure}
\centering
\includegraphics[width=0.8\textwidth]{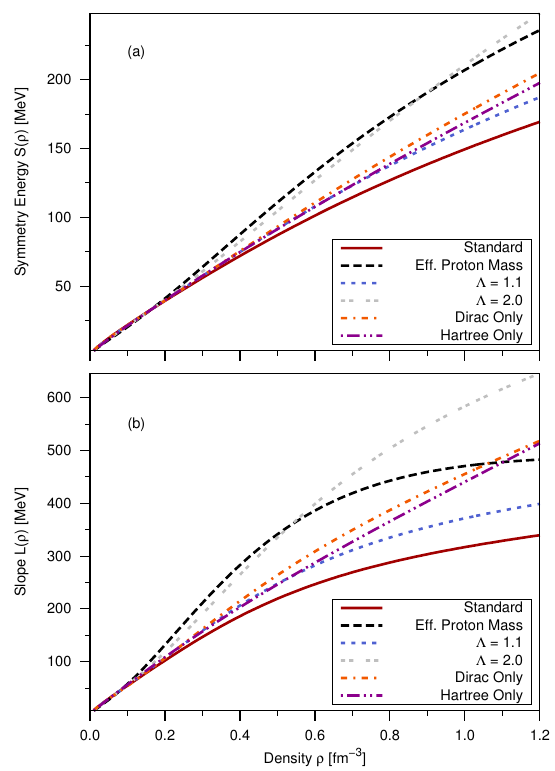} 
\caption{(Color online) (a) Symmetry energy $\mathcal{S}$ as a function 
of baryon number density, as calculated in this work. (b) 
Slope $L$ of the symmetry energy,  
as a function of baryon number density 
$L(\rho )=3\rho\left( \frac{\partial \mathcal{S} }{ \partial\rho} \right)$.}
\label{fig:SymSlopePlots}
\end{figure}

\clearpage
\begin{figure}
\includegraphics[width=\textwidth]{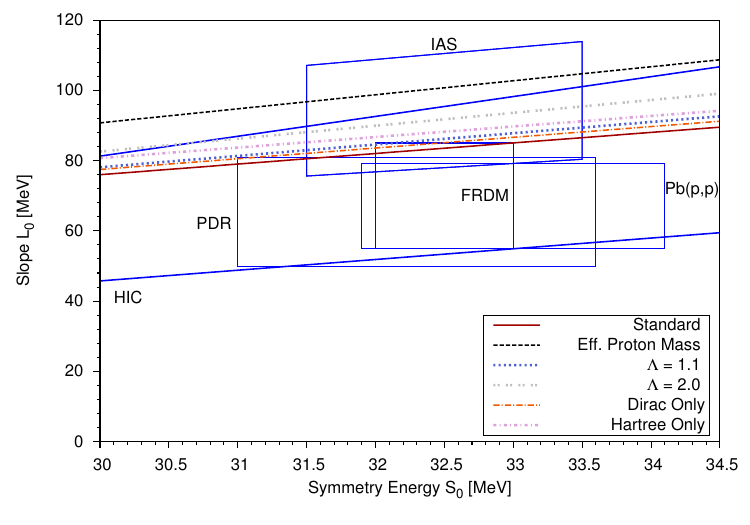} 
\caption{(Color online) The correlation between the slope and magnitude 
of the symmetry energy $\mathcal{S}_{\rm 0}$. 
Constraints on the slope $L_{0}$ and the symmetry energy
$S_0$ at saturation density from different
experiments are overlayed. The experimental methods are 
labeled next to the boxes
with the estimated uncertainties. See Ref.~\cite{Tsang:2012se} for more details.}
\label{fig:LinearSL}
\end{figure}

\clearpage
\begin{figure}
\includegraphics[width=\textwidth]{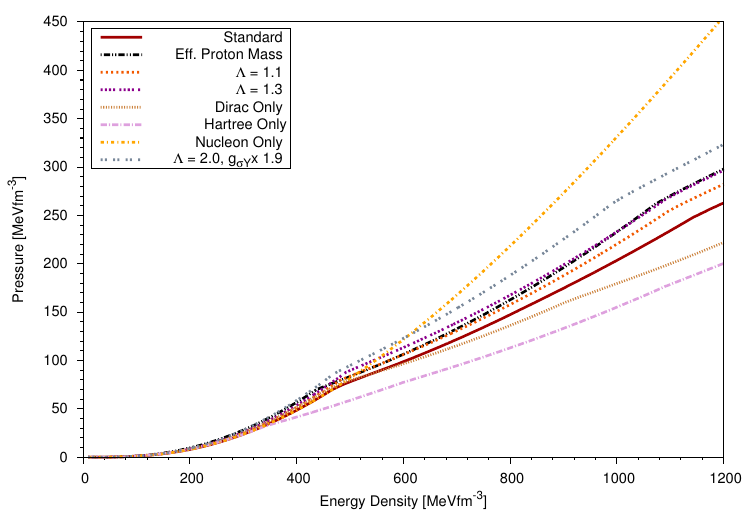} 
\caption{(Color online) GBEM equation of state. 
Kinks occur at significant hyperon threshold
densities. The divergences between the ``Hartree Only" QMC parameterization 
and the Hartree--Fock scenarios highlights the importance of Fock terms 
at high density. The ``Nucleon only'' 
BEM EoS is added for a comparison.
}
\label{fig:BEMEoS}
\end{figure}

\clearpage
\begin{figure}
\centering
\includegraphics[width=\textwidth]{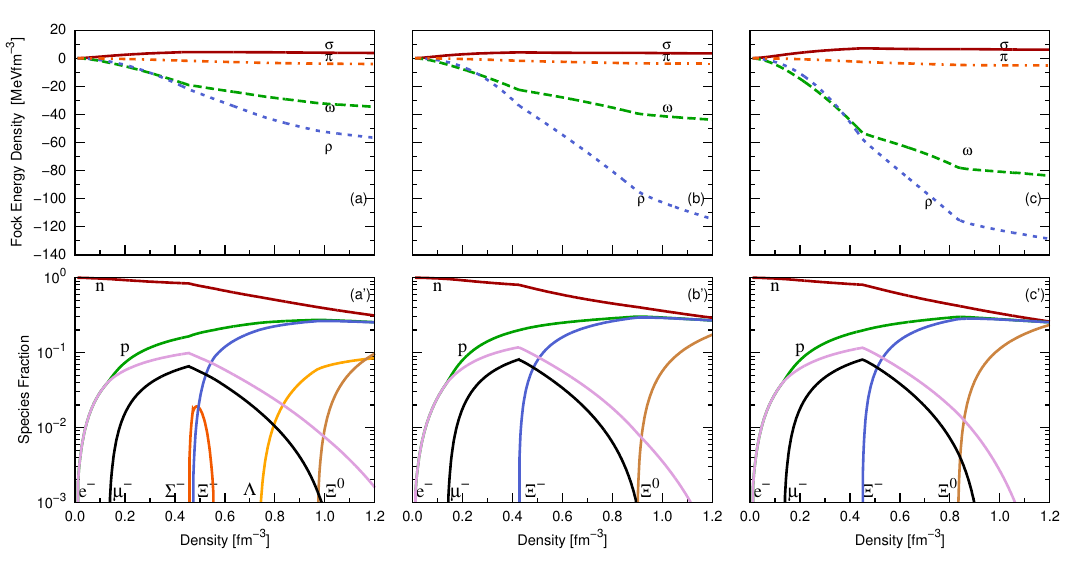}
\caption{(Color online) (Top) Fock energy density contributions and (bottom)  species fraction as a function of baryon number density in GBEM, 
for the ``Standard" (a,a'), ``Eff. Proton Mass" (b,b') and the ``$\Lambda=2.0$, $g_{\sigma Y}\times1.9$ (c,c') scenarios. The corresponding EoSs are shown in Fig.~\ref{fig:BEMEoS}}
\label{fig:SpFr}
\end{figure}

\clearpage
\begin{figure}
\centering
\includegraphics[width=0.8\textwidth]{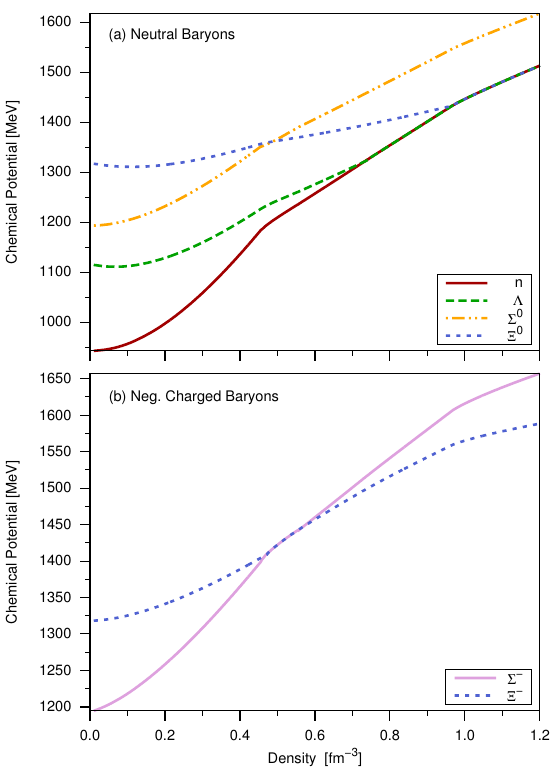}
\caption{(Color online) (a) Neutral baryon chemical potentials as a function of baryon number density
for the standard scenario. 
(b) Negative charge baryon chemical potentials as a function of baryon number density for the standard scenario.}
\label{fig:ChemPots}
\end{figure}

\clearpage
\begin{figure}
\includegraphics[width=\textwidth]{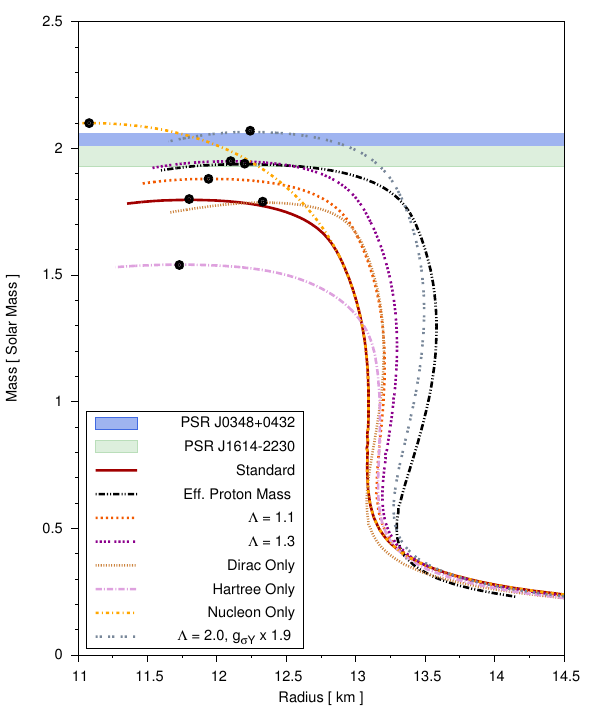} 
\caption{Gravitational Mass versus radius relationship for 
various scenarios described in the text. The black dots represent 
maximum mass stars and the coloured bars represent observed pulsar constraints.}
\label{fig:MVsR}
\end{figure}

\clearpage
\begin{figure}
\includegraphics[width=\textwidth]{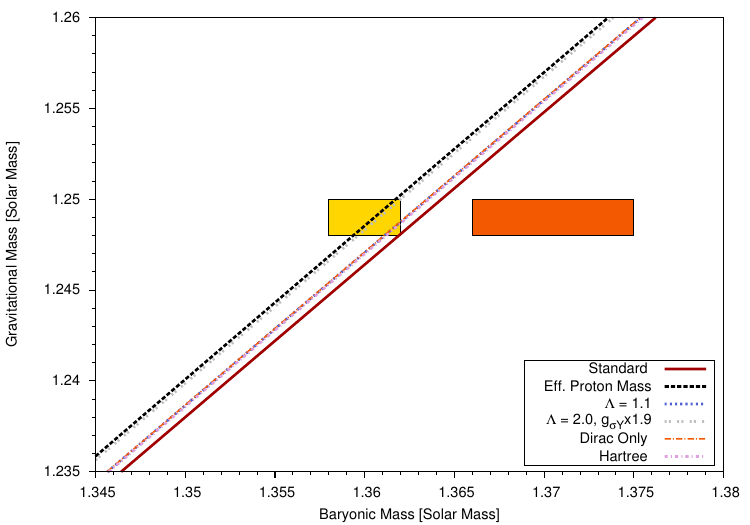} 
\caption{Gravitational mass versus baryonic mass.
The boxes are constraints from simulations (Yellow) 
by Kitaura {\it et al.}~\cite{kitaura2006}
and (Orange)  by Podsiadlowski {\it et al.}~\cite{podsiadlowski2005},
which are explained in the text. 
}
\label{fig:MgVsMb}
\end{figure}

\clearpage

\begin{table}[!htb]
\begin{center}
\begin{tabular}{|c|c|c|}
\hline
  &  &  \\
 Relation & Magnetic Moments [n.m.] & $\kappa^{B}_{(IS,IV)} =: \kappa^{B}_{(\omega,\rho)} $ 
\\
  &  &  \\
 \hline
  &  &  \\
$\mu_{p} =1 + \frac{1}{2}(\kappa^{N}_{IS}+\kappa^{N}_{IV})$ & $\mu_{n} = -1.913$& $\kappa^{N}_{IS} =-0.12 $\\
$\mu_{n} = \frac{1}{2}(\kappa^{N}_{IS}-\kappa^{N}_{IV})$ & $\mu_{p} = 2.793$&$\kappa^{N}_{IV} = 3.706$\\
  &  &  \\

 \hline
   &  &  \\

 $\mu_{\Lambda} = \kappa^{\Lambda}_{IS} $& $\mu_{\Lambda} = -0.61$&  $\kappa^{\Lambda}_{IS} =-0.61$ \\
   &  &  \\

 \hline
   &  &  \\

 $\mu_{\Sigma^{+}} =1 + (\kappa^{\Sigma}_{IS}+\kappa^{\Sigma}_{IV} )$ & $\mu_{\Sigma^{-}}=-1.16$& $\kappa^{\Sigma}_{IS}=0.649$ \\
 $ \mu_{\Sigma^{-}} =-1 + (\kappa^{\Sigma}_{IS}-\kappa^{\Sigma}_{IV})$ & $\mu_{\Sigma^{+}} =  2.458$ &  $\kappa^{\Sigma}_{IV}=0.809$\\
   &  &  \\

 \hline
   &  &  \\

 $\mu_{\Xi^{0}} = \frac{1}{2}(\kappa^{\Xi}_{IS}+\kappa^{\Xi}_{IV})$&$\mu_{\Xi^{-}} =-0.65$  &  $\kappa^{\Xi}_{IS} =-0.9$\\
 $\mu_{\Xi^{-}} = -1 + \frac{1}{2}(\kappa^{\Xi}_{IS}- \kappa^{\Xi}_{IV})$& $\mu_{\Xi^{0}} =-1.25$ &  $\kappa^{\Xi}_{IV}=-1.5993$\\
   &  &  \\
 \hline
\end{tabular}
\caption{Relations between baryon magnetic moments and anomalous isoscalar and isovector magnetic moments $\kappa^{B}_{(IS,IV)} =: \kappa^{B}_{(\omega,\rho)} = f_{B(\omega,\rho)}/g_{B(\omega,\rho)}$ using experimental magnetic moments \cite{Beringer:1900zz}. }
\label{table:kappas}
\end{center}
\end{table}

\singlespacing
\begin{sidewaystable}
    \centering
    \caption{ Couplings, nuclear matter properties, selected hyperon optical potentials
    and neutron star properties determined for our standard 
case (for which $\Lambda=0.9$~GeV, and 
$R^{\rm free}_{N}=1.0$~fm) and the effect of subsequent variations in which differences
from the standard parameter set are indicated in column 1. 
The tabulated quantities at saturation are the slope  and curvature of the symmetry
energy, $L_{0}$ and $K_{sym}$, the
incompressibility $K_{0}$, skewness coefficient $Q_{0}$, calculated at
saturaion density, and volume component of isospin 
incompressibility $K_{\tau, v}$, respectively.
Tabulated neutron star quantities are the stellar radius, maximum stellar mass and 
corresponding central density (units $\rho_0=0.16$~fm$^{-3}$).
\protect\label{table:couplings}}    
    \begin{small}    
\begin{tabular}{lcccccccccccccc}
\hline
  \multirow{2}{*}{Model/} & \multirow{2}{*}{\ $g_{\sigma N}$\ } &\multirow{2}{*}{\ $g_{\omega N}$\ }&\multirow{2}{*}{\ $g_{\rho}$\ }& $K_{0}$&$L_{0}$ & $K_{sym}$   & $Q_{0}$ & $K_{\tau, v}$& 
  $U_{\Lambda}$ & $U_{\Sigma^{-}}$ & $U_{\Xi^{-}}$& $M_{\rm max}$& $R$
  & $\rho^{\rm max}_{c}$\\
 Scenario&& & & (MeV)& (MeV) &(MeV)&(MeV)& (MeV) &(MeV)& (MeV) &(MeV)  & ($M_\odot$) & (km)& ($\rho_0$) \\[2mm]
 \hline\\
{\footnotesize Standard }  & 8.97 & 9.38 & 4.96 &273 & 84 & -23 & -305 & -431 & 3 & 26 & 5 & 1.80 & 11.80 & 5.88  \\   
{\footnotesize $\Lambda=1.0$ }   & 9.07 & 9.73 & 5.05 & 278 & 85& -15&-282 &-439 &10  &32  &8  & 1.84 & 11.86 & 5.82 \\ 
{\footnotesize $\Lambda = 1.1$ }   & 9.16 & 10.06 & 5.16 & 283 & 86 & -8 & -261 & -446 & 16  & 39  & 11 & 1.88 &11.94  &5.70  \\ 
{\footnotesize $\Lambda =1.2$ }   & 9.24 & 10.37 & 5.28  & 286 & 87 &-2 &-241 &-451 & 23 &46  & 15  & 1.92 &12.03  & 5.60 \\ 
{\footnotesize $\Lambda = 1.3$ }   & 9.31 & 10.67  & 5.40  &289 & 88 & 4 & -224 &-456 & 29  & 53  & 18  &  1.95& 12.10 & 5.52  \\ 
{\footnotesize $\Lambda =1.1$, $g_{\sigma Y}\times 1.3$ }   & 9.16 &
10.06 &5.16  &283 &86 &-8 &-261 &-446 & -15 &14  &-4  &1.84  &11.91 &5.78  \\ 
{\footnotesize $\Lambda =1.3$, $g_{\sigma Y}\times 1.3$}   &9.31 &10.67  & 5.40 &289 &88 &4 &-224 &-456 & -3 & 28 &3  &1.92  &12.01  &5.66  \\ 
{\footnotesize $\Lambda =2.0$, $g_{\sigma Y}\times 1.9$ }   &9.69
&12.27  &6.16  & 302&92 &31 &-137 &-478 &-29  &20  &-7  &2.07  &12.24&5.38  \\ 
{\footnotesize Increased $f_{\rho N}/g_{\rho N}$ }   &8.70 & 9.27  &3.86  & 267 & 81 & -34 & -321 &-424 &6  & 27  &6  &1.77  &11.61  &6.14  \\ 
{\footnotesize Fock $\delta\sigma$ }   &9.01 & 9.44  & 4.97  & 273 & 84 & -21 & -296 &-432  &4  &26  &5  & 1.81 & 11.82 & 5.86 \\ 
 {\footnotesize Eff. Proton Mass}  & 10.40 & 11.0  &4.55  &297 &101 &64 &-190 &-476 &11  &41  &10  & 1.94 &12.20  &5.48  \\ 
{\footnotesize Eff. Proton Mass, $\Lambda = 1.1$ }   & 11.08 & 12.31  & 4.85  & 311 & 111 & 126 & -87 &-509 & 34  & 67 & 22  &  2.07& 12.57 & 5.08  \\ 
{\footnotesize Eff. Proton mass $+ \ \delta\sigma$ }   &10.89 &11.55  &4.53  &285 &109 &132 &-232 &-432 &17  &49  &13  &1.99  &12.22  &5.46  \\ 
{\footnotesize Dirac Only}   & 10.10 & 9.22 & 7.84  & 294 & 85 & 0 &-299 & -424 &-23  & 4  &-8  &1.79  &12.33  &5.22  \\ 
{\footnotesize Hartree Only}   & 10.25 & 7.95 & 8.40  & 283 & 88 & -17 & -455& -405 & -49   & -23  & -21  & 1.54 &11.73  &6.04  \\ 
{\footnotesize Nucleon Only }    & 8.97 & 9.38 & 4.96 &273 & 84 & -23 & -305 & -431 & 3 & 26 & 5 & 2.10 & 11.08 &6.46  \\  
{\footnotesize $R=0.8$ }   & 9.30 & 9.85 & 4.98 & 277 & 85 & -15 & -269 & -443 & 6  & 25  & 5 &1.83  & 11.88 & 5.80 \\ 
{\footnotesize App. $S_{0}=32.5$ }   & 9.05 & 9.38 & 4.86 & 275 & 82 & -27 & -303 & -429 & 2 & 24 &  4& 1.80 &  11.82& 5.82 \\ 
{\footnotesize App. $S_{0}=30.0$ }   & 9.31 & 9.35  & 4.50 & 280 & 74 & -24 & -298 & -391 & -4  & 19  & 1  &  1.81& 11.82 &5.76  \\ 
{\footnotesize $S_{0} =30.0$ }   & 9.24 & 9.36 & 4.61 & 278 & 76 & -20 & -299 & -394& -2 &21  & 2  &1.81  & 11.81 &5.80  \\ 
\hline
    \end{tabular}
        \end{small}

\end{sidewaystable}


\clearpage
\begin{thebibliography}{99}
\vspace{-5mm}
\bibitem{dbhf} 
B. ter Haar, R. Malfliet, Phys. Rep. {\bf 149}, 207 (1987); 
E.N.E. van Dalen, C. Fuchs, A. Faessler, Nucl. Phys. {\bf A741}, 227 (2004);
F. de Jong, and H. Lenske, Phys. Rev.{\bf C57}, 3099 (1998).

\bibitem{bhf}
 J. Cugnon, P. Deneye, A. Lejeune, Z. Phys. {\bf A328}, 
409 (1987); W. Zuo, A. Lejeune, U. Lombardo, J. F. Mathiot, Eur. Phys. J.
{\bf A14}, 469 (2002); I. Bombaci, U. Lombardo, Phys. Rev. {\bf C44}, 1892
(1991); B. D. Day and R. B. Wiringa, Phys. Rev. {\bf C32}, 1057 (1985); 
M. Baldo, G. F. Burgio, and H. -J. Schulze, Phys. Rev. {\bf C61}, 055801
(2000); I. Vidana, A. Polls, A. Ramos, L. Engvik, and M. Hjorth-Jensen, Phys.
Rev. {\bf C62}, 035801 (2000).

\bibitem{variational}
A. Akmal, V. R. Pandharipande, D. G. Ravenhall, Phys. Rev. {\bf C58}, 1804
(1998); A. Mukherjee, V. R. Pandharipande, Phys. Rev. {\bf C75}, 035802 (2007).

\bibitem{cbf}
A.~Fabricioni and S.~Fantoni, Phys. Lett. {\bf B298}, 263 (1993); C. Bisconti,
F. Arias de Saavedra, G. Co', and A. Fabrocini, Phys. Rev. {\bf C73}, 054304
(2006).

\bibitem{dewulf2003} 
Y.~Dewulf, W.~H.~Dickhoff, D.~Van Neck, E.~R.~Stoddard and M.~Waroquier, 
Phys. Rev. Lett. {\bf 90}, 152501 (2003).

\bibitem{frick2003}
T.~Frick and H.~M\"{u}ther, Phys. Rev. {\bf C68}, 034310 (2003).

\bibitem{qmoca} B. S. Pudliner, V. R. Pandharipande, J. Carlson, S. C. Pieper,
and R. B. Wiringa, Phys. Rev. C 56, 1720 (1997); K. E. Schmidt and S. Fantoni, Phys. Lett. B 446, 99 (1999); 
J. Carlson, J. Morales Jr., V. R. Pandharipande, and D. G.
Ravenhall, Phys. Rev. C 68, 025802 (2003); S. Gandolfi, F. Pederiva, S. Fantoni, and K. E. Schmidt, Phys.
Rev. Lett. 99, 022507 (2007); S. Gandolfi, A. Yu. Illarionov, S. Fantoni, F. Pederiva, and
K. E. Schmidt, Phys. Rev. Lett. 101, 132501 (2008);  S. Gandolfi, A. Yu. Illarionov, K. E. Schmidt, F. Pederiva, and S. Fantoni, Phys. Rev. C 79, 054005 (2009). 

\bibitem{hebeler2010a}
K. Hebeler and A. Schwenk, Phys. Rev. C 82, 014314
(2010).
%
\bibitem{hebeler2010b}
 K. Hebeler, J. M. Lattimer, C. J. Pethick, and A. Schwenk,
Phys. Rev. Lett. 105, 161102 (2010).
%
\bibitem{Dutra:2012mb} 
  M.~Dutra, O.~Lourenco, J.~S.~Sa Martins, A.~Delfino, J.~R.~Stone and P.~D.~Stevenson,
  Phys.\ Rev.\ C {\bf 85}, 035201 (2012).
%
\bibitem{Guichon:1987jp} 
P.~A.~M.~Guichon,
Phys.\ Lett.\ B {\bf 200}, 235 (1988).  
%
\bibitem{Guichon:1995ue} 
P.~A.~M.~Guichon, K.~Saito, E.~N.~Rodionov and A.~W.~Thomas,
Nucl.\ Phys.\ A {\bf 601}, 349 (1996)  
%
\bibitem{Tsushima:1997rd} 
  K.~Tsushima, K.~Saito and A.~W.~Thomas,
  Phys.\ Lett.\ B {\bf 411}, 9 (1997)
  [Erratum-ibid.\ B {\bf 421}, 413 (1998)]
  [nucl-th/9701047].
%
\bibitem{Tsushima:1997cu} 
  K.~Tsushima, K.~Saito, J.~Haidenbauer and A.~W.~Thomas,
  Nucl.\ Phys.\ A {\bf 630}, 691 (1998)
  [nucl-th/9707022].
%
\bibitem{Guichon:2008zz} 
P.~A.~M.~Guichon, A.~W.~Thomas and K.~Tsushima,
Nucl.\ Phys.\ A {\bf 814}, 66 (2008).
%
\bibitem{Guichon:2004xg} 
P.~A.~M.~Guichon and A.~W.~Thomas,  
Phys.\ Rev.\ Lett.\  {\bf 93}, 132502 (2004)  
%
\bibitem{Guichon:2006er}
P.~A.~M.~Guichon, H.~H.~Matevosyan, N.~Sandulescu and A.~W.~Thomas,
Nucl.\ Phys.\ A {\bf 772}, 1 (2006)
%
\bibitem{Weissenborn:2011ut}
  S.~Weissenborn, D.~Chatterjee and J.~Schaffner-Bielich,
  Phys.\ Rev.\ C {\bf 85}, 065802 (2012)
%
\bibitem{Weissenborn:2011kb} 
  S.~Weissenborn, D.~Chatterjee and J.~Schaffner-Bielich,
  Nucl.\ Phys.\ A {\bf 881}, 62 (2012)

\bibitem{RikovskaStone:2006ta}
J.~Rikovska Stone, P.~A.~M.~Guichon, H.~H.~Matevosyan and A.~W.~Thomas,
Nucl.\ Phys.\ A {\bf 792} (2007) 341
%
\bibitem{deSwart:CGC} 
  J.~J.~de Swart,
  Rev.\ Mod.\ Phys.\  {\bf 35}, 916 (1963)
  [Erratum-ibid.\  {\bf 37}, 326 (1965)].

\bibitem{Krein:1998vc}
G.~Krein, A.~W.~Thomas and K.~Tsushima,
Nucl.\ Phys.\ A {\bf 650}, 313 (1999)
%
\bibitem{Miyatsu:2011bc} 
T.~Miyatsu, T.~Katayama and K.~Saito,
Phys.\ Lett.\ B {\bf 709}, 242 (2012)
%
\bibitem{Massot:2012pf}
E.~Massot, J.~Margueron and G.~Chanfray,
Europhys.\ Lett.\  {\bf 97}, 39002 (2012).
%
\bibitem{bissey2005}
F. Bissey, {\it et al.}, Nucl. Phys. B (Proc. Suppl.) {\bf 141}, 22 (2005) 
%
\bibitem{walecka}
B.~D.~Serot and J.~D.~Walecka, Adv. Nucl. Phys. 16, 1 (1986)
%
\bibitem{Bentz:2001vc} 
W.~Bentz and A.~W.~Thomas,
Nucl.\ Phys.\ A {\bf 696}, 138 (2001)
%
\bibitem{Glendenning:2000wn} 
N.~K.~Glendenning, Compact stars: Nuclear physics, particle physics, and general relativity,
New York, USA: Springer, second edition (2000) 
%
\bibitem{Williams:1996id}
  R.~A.~Williams and C.~Puckett-Truman,
  Phys.\ Rev.\ C {\bf 53} (1996) 1580.
%
\bibitem{Beringer:1900zz}
  J.~Beringer {\it et al.}  [Particle Data Group Collaboration],
  Phys.\ Rev.\ D {\bf 86} (2012) 010001.
%
\bibitem{Saito:2005rv}
  K.~Saito, K.~Tsushima and A.~W.~Thomas,
  Prog.\ Part.\ Nucl.\ Phys.\  {\bf 58} (2007) 1
%
\bibitem{blaizot1980}
J.~P.~Blaizot, Physics Reports, 64,171 (1980)

\bibitem{shlomo2006}
S. Shlomo, V.M. Kolomietz, and G. Colo, Eur. Phys. J. A 30, 23 (2006)

\bibitem{stone2013}
J.~R.~Stone, N.~J.~Stone and S.~A.Moszkowski, Phys. Rev. C89, 044316 (2014)

\bibitem{farine1997} 
M. Farine, J. M. Pearson, and F. Tondeur, Nucl. Phys. {\bf A615}, 135, (1997).

\bibitem{Tews:2012fj}
  I.~Tews, T.~Kruger, K.~Hebeler and A.~Schwenk,
  Phys.\ Rev.\ Lett.\  {\bf 110} (2013) 032504

\bibitem{hebeler2013}
K.~Hebeler and R.~J.~Furnstahl, Phys. Rev. C87, 031302R (2013)

\bibitem{horowitz2000}
C.~J.~Horowitz and J.~Piekarewicz, PRL 86, 5647 (2001)

\bibitem{vidana2009}
I.~Vidana, C.~Providencia, A.~Polls and A.~Rios, Phys. Rev. C80, 045806 (2009)

\bibitem{gandolfi2012}
S.~Gandolfi, J.~Carlson and S.~Reddy, Phys. Rev. C85, 032801¨ (2012)

\bibitem{Dutra:2013osa}
  M.~Dutra, O.~Lourenço, B.~V.~Carlson, A.~Delfino, D.~P.~Menezes, S.~S.~Avancini, J.~R.~Stone and C.~Providencia {\it et al.},Â
  arXiv:1303.2562 [nucl-th].
%
\bibitem{Sagawa:2007sp}
  H.~Sagawa, S.~Yoshida, G.~-M.~Zeng, J.~-Z.~Gu and X.~-Z.~Zhang,
  Phys.\ Rev.\ C {\bf 76} (2007) 034327
   [Erratum-ibid.\ C {\bf 77} (2008) 049902]
%
\bibitem{Tsang:2012se}
  M.~B.~Tsang, J.~R.~Stone, F.~Camera, P.~Danielewicz, S.~Gandolfi, K.~Hebeler, C.~J.~Horowitz and J.~Lee {\it et al.},
  Phys.\ Rev.\ C {\bf 86} (2012) 015803
%
\bibitem{danielewicz2002}
P.~Danielewicz, R.~Lacey and W.~G.~Lynch, Science 298, 1592 (2002)
%
\bibitem{stone2007a}
J.~R.~Stone and P.-G.~Reinhard, Prog. Part. Nucl. Phys. 58, 587 (2007)
%
\bibitem{Gandolfi:2011xu}
  S.~Gandolfi, J.~Carlson and S.~Reddy,
  Phys.\ Rev.\ C {\bf 85} (2012) 032801
%
\bibitem{treiner1981}
J.~Treiner, H.~Krivine, O.~Bohigas and J.~Martorell, Nucl. Phys. A 371, 253 (1981)
%
\bibitem{nayak1990}
R.~C.~Nayak, J.~M.~Pearson, M.~Farine, P.~Gleissl and M.~Brack, Nucl. Phys. A516, 62 (1990
%
\bibitem{Carroll:2010ex}
  J.~D.~Carroll,
  AIP Conf.\ Proc.\  {\bf 1374} (2011) 205
%
\bibitem{tov}
R.~C.~Tolman. Proc. Natl. Acad. Sci. U.S.A. 20, 3 (1934); J.~R.~Oppenheimer and G.~M.~Volkov, 
Phys. Rev. 55, 374 (1939)
%
\bibitem{Baym:1971pw} 
  G.~Baym, C.~Pethick and P.~Sutherland,
  Astrophys.\ J.\  {\bf 170}, 299 (1971).
%
\bibitem{Demorest:2010bx}
P.~Demorest {\it et al.}, 
Nature {\bf 467} (2010) 1081.
%
\bibitem{Antoniadis:2013pzd} 
  J.~Antoniadis, P.~C.~C.~Freire, N.~Wex, T.~M.~Tauris, R.~S.~Lynch, M.~H.~van Kerkwijk, M.~Kramer and C.~Bassa {\it et al.},
  Science {\bf 340}, 6131 (2013)
  [arXiv:1304.6875 [astro-ph.HE]].

%
\bibitem{steiner2010}
A.~W.~Steiner, J.~M.~Lattimer and E.~F.~Brown, ApJ 722, 33 (2010)
%
\bibitem{guillot2013}
S.~Guillot, M.~Servillat, N.~A.~Webb and R.~E.~Rutledge, arXiv:1302.0023 (2013)
%
\bibitem{podsiadlowski2005}
Ph.~Podsiadlowski, J.~D.~M.~Devi. P.~Lasaffre, J.~C.~Miller, W.~G.~Newton and J.~R.~Stone, Mon.Not.R.Astron.Soc. 361, 1243 (2005)
%
\bibitem{kitaura2006}
F.~S.~Kitaura, H.-Th.~Janka and W.~Hillebrant, Astron. Astrophys. 450, 345 (2006)
%

\bibitem{Rijken:2010zzb}
T.~A.~Rijken, M.~M.~Nagels and Y.~Yamamoto,
Prog.\ Theor.\ Phys.\ Suppl.\  {\bf 185}, 14 (2010).
%
\bibitem{SchaffnerBielich:2010am}
J.~Schaffner-Bielich,
Nucl.\ Phys.\  A {\bf 835}, 279 (2010)
%
\bibitem{Weber:2005}
F.~Weber, 
Prog.~Part.~Nucl.~Phys.~{\bf 54}, 193 (2005)
%
\bibitem{Miyatsu:2012xh}
  T.~Miyatsu and K.~Saito,
  arXiv:1209.3360 [nucl-th].
%
\bibitem{Katayama:2012ge}
  T.~Katayama, T.~Miyatsu and K.~Saito,
  Astrophys.\ J.\ Suppl.\  {\bf 203} (2012) 22
%
\bibitem{Miyatsu:2010zz}
  T.~Miyatsu and K.~Saito,
  Prog.\ Theor.\ Phys.\  {\bf 122} (2010) 1035.
%
\bibitem{Bouyssy:1987sh} 
  A.~Bouyssy, J.~F.~Mathiot, N.~Van Giai and S.~Marcos,
  Phys.\ Rev.\ C {\bf 36}, 380 (1987).
\bibitem{Chen:2009wv} 
  L.~-W.~Chen, B.~-J.~Cai, C.~M.~Ko, B.~-A.~Li, C.~Shen and J.~Xu,
  Phys.\ Rev.\ C {\bf 80}, 014322 (2009)
 .
  
\bibitem{Masuda:2012ed} 
  K.~Masuda, T.~Hatsuda and T.~Takatsuka,
  PTEP {\bf 2013}, no. 7, 073D01 (2013)
  [arXiv:1212.6803 [nucl-th]].
  
\bibitem{Masuda:2012kf} 
  K.~Masuda, T.~Hatsuda and T.~Takatsuka,
  Astrophys.\ J.\  {\bf 764}, 12 (2013)
  [arXiv:1205.3621 [nucl-th]].
  
  
  
\bibitem{Blaschke:2013rma} 
  D.~Blaschke, D.~E.~Alvarez Castillo, S.~Benic, G.~Contrera and R.~Lastowiecki,
  PoS ConfinementX {\bf }, 249 (2012)
  [arXiv:1302.6275 [hep-ph]].
\end{thebibliography}
\end{document}